\documentclass[
amsmath,
amssymb,
aps, 
pra,
twocolumn, 
groupedaddress,
nobalancelastpage,
floatfix]
{revtex4-2}
\usepackage{mathtools}
\usepackage[colorlinks]{hyperref}
\usepackage{lipsum}
\usepackage{bbold}
\usepackage{bm}
\usepackage{dsfont}
\usepackage{enumerate}
\usepackage{graphicx}
\usepackage{float}
\usepackage {xcolor}
\usepackage{cleveref}
\usepackage{amsmath}

\usepackage{lipsum}

\DeclareMathOperator{\Tr}{Tr}

\renewcommand{\vec}[1]{\bm{#1}}	

\DeclarePairedDelimiterX{\mean}[1]{\langle}{\rangle}{
	{#1}
}
\DeclarePairedDelimiterX{\abs}[1]{\lvert}{\rvert}{
	{#1}
}
\DeclarePairedDelimiterX{\norm}[1]{\lVert}{\rVert}{
	{#1}
}
\DeclarePairedDelimiterX{\bra}[1]{\langle}{\rvert}{#1}

\DeclarePairedDelimiterX{\ket}[1]{\lvert}{\rangle}{#1}

\DeclarePairedDelimiterX{\mel}[3]{\langle}{\rangle}{
{#1}\delimsize\vert {#2}\delimsize\vert {#3}
}
\DeclarePairedDelimiterX{\inner}[2]{\langle}{\rangle}{
{#1} \delimsize\vert{#2}
}
\DeclarePairedDelimiterX{\dyad}[2]{\lvert}{\vert}{
	{#1} \delimsize\rangle \delimsize \langle{#2}
}

\begin{document}
	\title{High-Harmonic Generation in a Crystal Driven by Quantum Light}
	
	\author{Rasmus Vesterager Gothelf}
	\author{Christian Saugbjerg Lange} 
	\author{Lars Bojer Madsen}
	\affiliation{Department of Physics and Astronomy, Aarhus University, Ny Munkegade 120, DK-8000 Aarhus C, Denmark}
	\date{\today}
	\begin{abstract}
		We study intraband high-harmonic generation (HHG) in a crystal driven by quantum light. Previous theoretical studies have developed a framework based on coherent state expansions in terms of $P$ distributions to consider nonclassical driving fields for HHG in atoms. Here, we adapt this framework to the context of solids and consider an intraband model of ZnO.  We investigate the effect of the quantum optical nature of the driving field on the harmonic spectra including the cutoff and the intensity scaling of the harmonics with driving field intensity. Based on analytical calculations in the Floquet limit, we explain why driving with thermal light or bright-squeezed vacuum (BSV) produces a much higher cutoff than when driving with fields described by coherent or Fock states. Further, we derive an expression for the generated time-dependent electric field and its fluctuations and find that it inherits characteristics of the driving field. Finally, we discuss the limitations of an approximative positive $P$ representation, which is introduced to be able to reduce the numerical complexity for Fock and BSV driving fields.
    \end{abstract}
	
	\maketitle
	
	\newpage 
    
\section{Introduction}\label{Sec: introduction}
	High-harmonic generation (HHG), the non-linear upconversion of a low-frequency intense laser field via an electronic medium, has through the years proven to be a subject of much interest. After the phenomenon was observed \cite{Ferray1988} and analyzed theoretically \cite{Corkum1993,Lewenstein1994} it was used to produce sub-femtosecond laser pulses \cite{Paul2001,Hentschel2001}, which brought with them the research area of attosecond physics \cite{Krausz2009,Villeneuve2018,Vampa2017}.
	\par
	The original studies of HHG considered atomic gasses as the generating medium. 
	However, a new avenue for exploration opened when HHG with near-infrared fields was observed from a bulk crystal \cite{Ghimire2011}. 
	HHG from solids proved to have many unique characteristics,
	such as the harmonic cutoff scaling linearly with field strength \cite{Ghimire2011}, as opposed to the quadratic scaling in gasses, as well as, e.g., topological features that affect the spectra \cite{Bauer2018a,Bai2021,Morimoto2016,Chacon2020,Silva2019}. 
	Understanding of HHG in this context is reached through the three-step model \cite{Ghimire2012,Vampa2014,Vampa2017,Yue2022}. First, electrons are promoted from the valence band to a conduction band. In the conduction band, the electrons propagate driven by the external electric field. Lastly, the electrons recombine with their holes in the valence band, and the excess energy is released as light. Because of the nonparabolic shape of the conduction band, high harmonics are also generated in the propagation step. As such, it has been useful to divide theoretical considerations of HHG in solids into generation from intraband and interband currents.
	Not only was this development a subject of scientific curiosity, it also pointed towards applications, such as using optical methods to reconstruct the conduction bands \cite{Vampa2015}. 
	\par
	In these investigations, the electromagnetic field used to drive HHG as well as the generated field were modeled by classical fields, while the electronic degrees of freedom were typically treated quantum mechanically. This approach has produced many successful results, as the quantum properties of intense pulses of  coherent light are typically negligible when considering intense-field-induced processes.
	\par
	Recently, a shift towards a fully quantum description of HHG is developing. This is motivated by the wish to combine attoscience and quantum optics, involving the investigation of the quantum nature of the emitted light. These quantum degrees of freedom of the emitted light can be used as spectroscopic measures to study the HHG medium and potentially create attosecond pulses of nonclassical light in the XUV regime. Overall, this field of study can be categorized into two areas: (1) fully quantum descriptions of HHG  when driven by coherent light \cite{Gorlach2020,Stammer2022,Stammer2023,Bhattacharya2023,Lange2024a, Theidel2024, Theidel2024b, Lange2024b,Yi2024}, and (2) HHG driven by quantum light \cite{Gorlach2023,EvenTzur2023,Tzur2024,Stammer2024}, such as squeezed light or Fock states. In both cases, the accurate treatment of the quantum degrees of freedom of both electrons \textit{and} photons presents a challenge to theory.
	\par 
	In the first category, the theoretical framework for considering quantum fields in the context of HHG is described in detail in, e.g., Refs. \cite{Gorlach2020,Stammer2023}. 
    This methodology has been used to show that the harmonic modes are entangled \cite{Gorlach2020,Stammer2022,Lange2024a,Yi2024}, that electron correlations in many-body systems can cause non-classicality in the produced light \cite{Lange2024a,Pizzi2023}, that HHG can be used to generate optical cat states \cite{Lewenstein2021}, and that a so-called Markov-state approximation, where the photonic degrees of freedom are accounted for in a closed analytical form can be accurate \cite{Lange2024b}. Recently the quantum optical nature of HHG was verified experimentally in crystals \cite{Theidel2024, Theidel2024b}.
	\par
	In the second category, models of HHG in atomic gasses driven with quantum light have been presented, which use coherent-state expansions in terms of generalized $P$ distributions to describe the photonic degrees of freedom \cite{Gorlach2023,EvenTzur2023}. This coherent-state expansion is attractive, as it allows for theory and methods developed for a semiclassical description to be applied. The theoretical results show that quantum properties of the driving field can affect characteristics of the harmonic spectrum such as the cutoff \cite{Gorlach2023}. Furthermore, it has been shown that driving HHG with squeezed coherent light produces squeezed harmonics \cite{Tzur2024}, indicating that quantum properties of the driving field are mapped onto the properties of the generated field. To this end, we note that new experimental methods allow for the production of nonclassical bright-squeezed vacuum (BSV) fields, which are becoming sufficiently intense to drive HHG \cite{Rasputnyi2024,Manceau2019}. Such a BSV field can also be used to perturb a stronger coherent driving field resulting in the emission of bunched sidebands from HHG \cite{Lemieux2024}.
	\par 
	As a whole, the quantum-optical consideration of HHG has multiple prospects. Besides the ability to optimize the cutoff, the study bridges the gap between strong–field physics and quantum optics. HHG has the potential to produce and control high-intensity quantum states of light, which points towards applications in the field of quantum information \cite{Bhattacharya2023,Stammer2022,Rivera-Dean2024c,Braunstein2005}. 
	\par
	As stated, most theoretical models of HHG driven by quantum light have so far considered the medium of atomic gasses \cite{Gorlach2023,EvenTzur2023}. 
    This paper aims to extend this development to HHG in crystals driven by quantum light. To do so, we adapt the newly developed models for non-classical driving of HHG \cite{Gorlach2023, EvenTzur2023} to the established model for intraband HHG in crystals. The intraband HHG description is sufficiently simple to lend itself to analytic considerations, which we use to determine and interpret the characteristics of the generated spectrum. Hereby, we show that these recent developments can also be applied to solids, and we identify some of the immediate differences between HHG in gasses and crystals.
	\par
    The paper is organized as follows. In Sec. \ref{Sec: Theory} we present the derivation of the generated photonic state for intraband HHG driven by quantum light, and we discuss the different choices of coherent state expansions of the driving fields, i.e., the $P$ representations that are considered. 
    In Sec. \ref{Sec: harmonic_spectrum} we discuss the HHG spectrum. We consider the Floquet limit of a time-periodic driving field and relate these results to simulations of the generated field using a 16-cycle pulse. From this, we study the dependence of the harmonic cutoff of the spectra on the quantum optical nature of the driving field and the scaling of the harmonic intensity in terms of the driving field intensity. In Sec. \ref{Sec: E-field} we derive an expression for and compute the time-resolved electric field generated from the HHG spectrum. In Sec. \ref{Sec: Accuracy_of_Q} we discuss the accuracy of an introduced approximative positive $P$ representation that is considered in the case of Fock and BSV driving fields. 
    Lastly, in Sec. \ref{Sec: conclusion} we give a conclusion and an outlook. The Appendices \ref{App: squeez_in_Q}, \ref{App: Bessel_approx}, and \ref{App: E_field} give details on the accuracy of describing different observables with an approximate positive $P$ representation, the accuracy of the lowest-order expansion of the Bessel function involved in the analysis of the perturbative regime, and the derivation of the generated time-dependent electric field, respectively. 
    \par
    Atomic units ($\hbar=m_e=4\pi\epsilon_0=e=1$) are used throughout the paper unless stated otherwise.

\section{Theory}\label{Sec: Theory}

The theory of HHG in solids driven by quantum light is based on the models of Refs. \cite{Gorlach2023,EvenTzur2023} developed for atoms. For completeness, we present an overview of the derivation of the generated field. Readers familiar with these matters may therefore skip this section. We will be utilizing the velocity gauge as in Refs. \cite{Lange2024a, Lange2024b} and as opposed to Refs. \cite{Gorlach2023,EvenTzur2023}, since this is a more convenient gauge for considering solids, as the interaction is independent of electron spatial coordinate in the dipole approximation, and Bloch's theorem therefore still applies. We will be deriving the model using a general $P$ representation to describe the photonic degrees of freedom. As there are multiple choices of such representations, we keep the notation general for the derivation of the generated field and specify the choice of representation later (Sec. \ref{Sec: Representation}).

\subsection{Derivation of the photonic state}
	Consider the minimally coupled, fully quantized Hamiltonian in the velocity gauge
	\begin{align}
		\hat{H} = \frac{1}{2}\sum_j \left( \hat{\vec{p}}_j + \hat{\vec{A}}\right)^2 + \hat{V}_e + \hat{H}_F,
		\label{Eq: Hamiltonian}
	\end{align}
	where $\hat{\vec{p}}_j$ is the momentum of the $j$'th electron, $\hat{V}_e$ is the electronic potential, $\hat{H}_F=\sum_{\vec{k},\sigma}\omega_{\vec{k}} \hat{a}_{\vec{k},\sigma}^\dag \hat{a}_{\vec{k},\sigma}$ 
    is the free-field Hamiltonian with $ \hat{a}_{\vec{k},\sigma}$ 
    ($\hat{a}^\dag_{\vec{k},\sigma}$) being the annihilation (creation) operator of the field mode of wave vector $\vec{k}$ and polarization $\sigma$ and $\omega_{\vec{k}}$ being the frequency of the corresponding mode, and lastly
	$\hat{\vec{A}} = \sum_{\vec{k},\sigma} \frac{g_0}{\sqrt{\omega_{\vec{k}}}}
	\left( \vec{e}_\sigma \hat{a}_{\vec{k},\sigma} + h.c.\right)$
	is the quantized vector potential in the dipole approximation with $\vec{e}_\sigma$ being the unit vector of the polarization, $g_0=\sqrt{2\pi/V}$ the coupling constant, and $V$ the quantization volume.
	\par 
	Similar to Refs. \cite{Gorlach2023,EvenTzur2023}, the state of the combined electronic and photonic system at the initial time $t_i$ is assumed to be described by the total density operator
	\begin{align}
		\hat{\rho}(t_i) = \dyad{\phi_i}{\phi_i}
		&\bigotimes_{(\vec{k},\sigma)\neq (\vec{k}_0,\sigma_0)}
		 \dyad{0_{\vec{k},\sigma}}{0_{\vec{k},\sigma}}
		 \notag
		 \\
		 &\otimes
		 \int d\mu P(\alpha,\beta)\frac{\dyad{\alpha}{\beta^*}}{\inner{\beta^*}{\alpha}}, \label{Eq: initial_state}
	\end{align}
	where $\ket{\phi_i}$ is the electronic ground state, $\ket{0_{\vec{k},\sigma}}$ is the vacuum state in the photonic mode of wave vector $\vec{k}$ and polarization $\sigma$. The driving field is assumed to be described by a single mode ($\vec{k}_0,\sigma_0$) and is expressed in terms of a generalized $P$ representation \cite{Drummond1980,Walls2008} over the coherent states of that mode. In this way, the theory is built in a general representation, and a specific representation can be chosen by specifying the measure $d\mu$ in Eq. (\ref{Eq: initial_state}) over the coherent phase space (Sec. \ref{Sec: Representation}). 
	
	To determine the time evolution of Eq. (\ref{Eq: initial_state}) through the Hamiltonian [Eq. (\ref{Eq: Hamiltonian})], the Ansatz is made that the time-dependent density matrix can be written as 
	\begin{equation}
		\hat{\rho}(t) = \int d\mu
		\frac{P(\alpha,\beta)}{\inner{\beta^*}{\alpha}}
		\hat{\rho}_{\alpha\beta^*}(t),
		\label{Eq: rho_ab_integral}
	\end{equation}
	where integration is still over the coherent phase space of the driving mode with the measure $d\mu$, and where $\hat{\rho}_{\alpha\beta^*}(t)$ is the time evolution of the state $\dyad{\phi_i}{\phi_i}\otimes
	\dyad{\alpha}{\beta^*}
	\bigotimes_{(\vec{k},\sigma)\neq (\vec{k}_0,\sigma_0)}
	\dyad{0_{\vec{k},\sigma}}{0_{\vec{k},\sigma}}
	$. By the linearity of the integral in Eq. (\ref{Eq: rho_ab_integral}), it is then apparent that such a solution would also determine the time evolution of Eq. (\ref{Eq: initial_state}).
	\par 
	To find $\hat{\rho}_{\alpha\beta^*}(t)$, the interaction picture with respect to $\hat{H}_F$ is employed.  
    This picture introduces the time-dependent phases $e^{\pm i\omega_{\vec{k}} t}$ to the vector potential 
    \begin{align}
        \hat{\vec{A}}(t) = \sum_{\vec{k},\sigma} \frac{g_0}{\sqrt{\omega_{\vec{k}}}}
	\left( \vec{e}_\sigma \hat{a}_{\vec{k},\sigma} e^{-i\omega_{\vec{k}} t} + h.c.\right).
	\label{eq:A_definition_dipole}
    \end{align} 
	\par 
	Next, another unitary transformation is applied, using displacement operators with the coherent state amplitudes $\alpha$ and $\beta^*$, such that the driving mode of $\hat{\rho}_{\alpha\beta^*}(t)$ is shifted to a vacuum state at time $t_i$, that is, we let $\hat{\tilde{\rho}}_{\alpha\beta}(t) = \hat{D}^\dag(\alpha) \hat{\rho}_{\alpha\beta}(t) \hat{D}(\beta^*)$. Applying this transform to the Hamiltonian in Eq. (\ref{Eq: Hamiltonian}), the vector potential transforms as $\hat{\vec{A}}(t)\mapsto \hat{\vec{A}}_Q(t) + \vec{A}^\alpha_{cl}(t)$ \cite{Cohen_Tannoudji1998_atomphoton}, where $ \hat{\vec{A}}_Q(t)$ is the same quantum vector potential as $\hat{\vec{A}}(t)$ in Eq. (\ref{eq:A_definition_dipole}) but now given a subscript for the sake of distinction, and where 
    \begin{align}
        \vec{A}_{cl}^\alpha(t) = \mel{\alpha}{\hat{\vec{A}}(t)}{\alpha}= \frac{g_0}{\sqrt{\omega_0}}
	\left( \vec{e}_{\sigma_0} \alpha e^{-i\omega_0 t} + h.c.\right)
    \label{Eq: Classical_vector_potential}
    \end{align}
    is the classical vector potential of the driving field corresponding to the coherent state $\ket{\alpha}$. 
    It then follows that
	\begin{align}
		i\frac{\partial \hat{\tilde{\rho}}_{\alpha\beta^* }(t)}{\partial t}
		=
		\hat{H}_{\alpha}(t) \hat{\tilde{\rho}}_{\alpha\beta^*}(t) - \hat{\tilde{\rho}}_{\alpha\beta^*}(t) \hat{H}_{\beta}(t),
		\label{Eq: rho_tilde_sl}
	\end{align}
	where 
	$
	\hat{H}_\xi(t) = \frac{1}{2}\sum_j \left( \hat{\vec{p}}_j + \hat{\vec{A}}_{Q}(t) + \vec{A}_{cl}^\xi(t)    \right)^2 + \hat{V}_e 
	$ for $\xi = \{\alpha,\beta\}$.
	We neglect $\hat{\vec{A}}_Q^2(t)$ as done in Refs. \cite{Gorlach2020,Lange2024a}. Hence,
	\begin{align}
		\hat{H}_\xi (t) =	\hat{H}_{SC}(t;\xi) + \hat{\vec{A}}_Q(t) \cdot \sum_j \left[ \hat{\vec{p}}_j + \vec{A}_{cl}^\xi(t)    \right],
	\end{align}
	where
	\begin{align}
		\hat{H}_{SC}(t;\xi) = \frac{1}{2}\sum_j \left[ \hat{\vec{p}}_j + \vec{A}_{cl}^\xi(t)    \right]^2 + \hat{V}_e,
		\label{Eq: semiclassical_H}
	\end{align}
    is the semiclassical (SC) Hamiltonian of the single-mode coherent field with amplitude $\xi$.
	\par 
    Letting $\ket{\Psi_\xi (t)}$ be the solution to the time-dependent Schrödinger equation \begin{align}
        i\frac{\partial}{\partial t} \ket{\Psi_\xi (t)} = \hat{H}_\xi (t) \ket{\Psi_\xi (t)}
        \label{Eq: psi_alpha_ligning_ikke_interaction}
    \end{align}
    with $\ket{\Psi_\xi (t_i)} = \ket{\phi_i} \bigotimes_{\vec{k},\sigma} \ket{0_{\vec{k},\sigma}}$, 
    it is apparent that $\dyad{\Psi_\alpha(t)}{\Psi_{\beta^*}(t)}$ is a solution to Eq. (\ref{Eq: rho_tilde_sl}) with the same initial conditions as $\hat{\tilde{\rho}}_{\alpha\beta^*}(t)$, and therefore $\hat{\tilde{\rho}}_{\alpha\beta^*}(t) = \dyad{\Psi_\alpha(t)}{\Psi_{\beta^*}(t)}$.
	\par 
	To determine $\ket{\Psi_\xi (t)}$, an interaction picture is employed via the time-evolution operator $\hat{\mathcal{U}}_{SC}(t; \xi)$ associated with $\hat{H}_{SC}(t;\xi)$. In this picture, Eq. (\ref{Eq: psi_alpha_ligning_ikke_interaction}) becomes
	\begin{align}
		i \frac{\partial}{\partial t} \ket{\Psi_\xi(t)} = \hat{\vec{A}}_Q(t)\cdot \bigg[\hat{\mathcal{U}}^\dagger_{SC}(t; \xi)\hat{\vec{j}}^\xi (t)\hat{\mathcal{U}}_{SC}(t; \xi)\bigg] \ket{\Psi_\xi(t)},
		\label{Eq: psi_alpha_ligning}
	\end{align}
	where $\hat{\vec{j}}^\xi(t) = \sum_j\left[ \hat{\vec{p}}_j + \vec{A}_{cl}^\xi(t)    \right]$ is the current operator.
	\par 
	Equation (\ref{Eq: psi_alpha_ligning}) may be solved by expanding in the electronic basis $\{\ket{\phi_m}\}$ of solutions to the field-free time-independent Schrödinger equation. Letting $\{\ket{\chi^\xi_m(t)} \}$ denote the corresponding photonic states, the complete state of the combined system can then be expressed as $\ket{\Psi_\xi(t)} = \sum_m \ket{\phi_m}\otimes\ket{\chi^\xi_m(t)}$. Projecting onto the $m$'th electronic state, $\bra{\phi_m}$, the equation of motion for the corresponding photonic state $\ket{\chi^\xi_m(t)}$ becomes
	\begin{align}
		i \frac{\partial}{\partial t}\ket{\chi^\xi_m(t)} =\hat{\vec{A}}_Q(t)\cdot  \sum_n  \vec{j}^\xi_{m,n}(t) \ket{\chi^\xi_n(t)},
		\label{Eq: photonic_sl}
	\end{align}
    where 
    \begin{align}
    	 \vec{j}^\xi_{m,n}(t) = \bra{\phi_m^\xi(t)} \hat{\vec{j}}^\xi(t) \ket{\phi_n^\xi(t)} 
    	 \label{Eq: current_matrix_element}
  	 \end{align}
    is the current matrix elements between different semiclassically propagated electronic states, $\ket{\phi^\xi_n(t)}=\hat{\mathcal{U}}_{SC}(t;\xi)\ket{\phi_n}$. We note that in the SC theory of HHG, the spectrum is related to the norm square of the Fourier transform of the current matrix element $ \vec{j}^\xi_{i,i}(t) = \bra{\phi_i^\xi(t)} \hat{\vec{j}}^\xi(t) \ket{\phi_i^\xi(t)}$ for $i$ denoting the initial state of the electronic problem. It is the presence of the non-diagonal transition current matrix elements in Eq. (\ref{Eq: photonic_sl}) that induce quantum light in the HHG process \cite{Gorlach2020, Lange2024a, Lange2024b,Yi2024}.
    
    Before we specify our solid-state model, a few remarks regarding the characteristics of the above developments are in place. First, we note that the expansion in coherent states in terms of the $P$ function allows the usage of SC theory and associated numerical techniques to solve the time-dependent Schrödinger equation. Secondly, the additional use of displacement operators to transform the initial state to be in the vacuum state allows one to handle the large number of photons considered in this problem, i.e., one only needs to numerically handle the photons generated from the HHG process which are much fewer than the ones in the driving field. These two ingredients are essential for the making the problem tractable numerically.

	We now specify the electronic system considered in this work. We use an intraband model which corresponds to considering a one-dimensional cut of the solid along the polarization direction of $\vec{A}^\xi_{cl}(t)$. We consider a lattice constant $a$ and beyond-nearest-neighbor terms, which in the crystal-momentum basis is expressed as 
	\begin{align}
			\hat{H}_{SC}(t;\xi) =\sum_{q,\mu}\mathcal{E}[q + A_{cl}^\xi (t)] ~ \hat{c}_{q,\mu}^\dag \hat{c}_{q,\mu} ,
			\label{Eq: tight_binding_H}
	\end{align}
	where $\hat{c}_{q,\mu}$ ($\hat{c}^\dag_{q,\mu}$) is the annihilation (creation) operator of the crystal momentum state $q$ with spin $\mu$, and where
		\begin{align}
	\mathcal{E}(q) = \sum_{l=0} b_l \cos (al q)
	\end{align}
	is the dispersion relation of the material, where $b_l$ is the $l$'th Fourier coefficient of the band structure. The current operator in this model is 
	\begin{align}
		\hat{\vec{j}}^\xi(t) =  {\vec{e}}_\sigma \sum_{q,\mu}
		\frac{\partial \mathcal{E}[q + A_{cl}^\xi (t)]}{\partial q}
		~\hat{c}_{q,\mu}^\dag \hat{c}_{q,\mu},
		\label{Eq: general_current}
	\end{align}
	which is a vector along the polarization direction. 
	\par
	From Eqs. (\ref{Eq: tight_binding_H}) and (\ref{Eq: general_current}) it is apparent that both $\hat{H}_{SC}(t;\xi)$ and $\hat{\vec{j}}^\xi(t)$ are diagonal in the basis of crystal-momentum states meaning that the off-diagonal terms in Eq. (\ref{Eq: current_matrix_element}) vanish.
    Hereby Eq. (\ref{Eq: psi_alpha_ligning}) reduces to 
    \begin{align}
		i\frac{\partial}{\partial t} \ket{\chi^\xi_m(t)}
		=
		\hat{\vec{A}}_Q \cdot \vec{j}^\xi_{m,m}(t) \ket{\chi^\xi_m(t)}.
		\label{Eq: photonic_sl_model_alle_m}
	\end{align}
  	We consider the case where the system is initially in its ground state, i.e., $\ket{\chi^\xi_m(t_i)} = \delta_{i, m} \ket{0}$, so we only need to consider the state with $m=i$. Equation (\ref{Eq: photonic_sl_model_alle_m}) thus reduces to 
	\begin{align}
		i\frac{\partial}{\partial t} \ket{\chi^\xi_i(t)}
		=
		\hat{\vec{A}}_Q \cdot \vec{j}^\xi_{i,i}(t) \ket{\chi^\xi_i(t)},
		\label{Eq: photonic_sl_model}
	\end{align}
	which is linear in creation and annihilation operators and can therefore be solved. The solution to Eq. (\ref{Eq: photonic_sl_model}) is a direct product of coherent states \cite{Gerry2004,Scully1997}
    \begin{align}
        \ket{\chi^\xi_i(t)} = \bigotimes_{\vec{k},\sigma}  \hat{D}[\gamma^\xi_{\vec{k},\sigma}(t)] \ket{0_{\vec{k},\sigma}},
        \label{Eq: chi_xi_solution}
    \end{align}
    over all considered photonic modes $({\vec{k},\sigma})$ including the laser mode, and where
	\begin{align}
		\gamma^\xi_{\vec{k},\sigma}(t) = 
		-i\frac{g_0}{\sqrt{\omega_{\vec{k}}}} 
		\int_{t_i}^t \vec{j}^\xi_{i,i}(t')\cdot {\vec{e}}_\sigma e^{i\omega_{\vec{k}} t'}dt'
		,
		\label{Eq: gamma}
	\end{align}
	is the coherent state amplitude which is the Fourier transform of the current matrix element in the Floquet limit of $t\rightarrow\infty$ and $t_i\rightarrow-\infty$. As such, the time dependence in the model is carried by these integral limits. The photonic state in Eq. (\ref{Eq: chi_xi_solution}) was also discussed in Ref. \cite{Lange2024a} and the result shows that only coherent light can be generated when driving a one-band model with coherent light.
	\par 
	The solution [Eq. (\ref{Eq: chi_xi_solution})] to Eq. (\ref{Eq: photonic_sl}) is substituted into Eq. (\ref{Eq: rho_ab_integral}) by writing $\hat{\tilde{\rho}}_{\alpha\beta^*}(t) = \dyad{\phi^\alpha_i(t)}{\phi^{\beta^*}_i(t)}\otimes\dyad{\chi_i^\alpha(t)}{\chi_i^{\beta^*}(t)}$. Hereby it is found that
    \begin{align}
		\hat{\rho}(t) =& 
		\int d\mu\frac{P(\alpha,\beta)}{\inner{\beta^*}{\alpha}}\dyad{\phi^\alpha_i(t)}{\phi^{\beta^*}_i(t)}
        \notag
		\\
        &\bigotimes
		\hat{D}(\alpha)
		\dyad{\gamma^\alpha_{k_0,\sigma_0}(t)}{\gamma^{\beta^*}_{k_0,\sigma_0}(t)}
		\hat{D}^\dag(\beta^*) \nonumber 
		\\
	    &\bigotimes_{(k,\sigma)\neq (k_0,\sigma_0)}
		\dyad{\gamma^\alpha_{k,\sigma}(t)}{\gamma^{\beta^*}_{k,\sigma}(t)}.
	\end{align}
	The SC evolution of the field-free eigenstates is simply given as $\ket{\phi^\xi_i(t)}= \exp[-i E_\xi(t)]\ket{\phi_i(t_i)}$, where $E_\xi(t) = \int_{t_i}^t dt'\sum_{q,\mu}
	\mathcal{E}[q + A_{cl}^\xi(t')] \mean{\hat{c}^\dag_{q,\mu}\hat{c}_{q,\mu}}$ \cite{Lange2024}. The electronic states are then traced out, leaving their phases, and the state of the field at time $t$ is given as
	\begin{align}
		\hat{\rho}_F(t) =& 
		\int d\mu\frac{P(\alpha,\beta)}{\inner{\beta^*}{\alpha}}
        e^{i[E_{\beta^*}(t)-E_\alpha(t)]}
        \notag
		\\
        &\bigotimes 
		\hat{D}(\alpha)
		\dyad{\gamma^\alpha_{k_0,\sigma_0}(t)}{\gamma^{\beta^*}_{k_0,\sigma_0}(t)}
		\hat{D}^\dagger(\beta^*)
		\nonumber
		\\
	    &\bigotimes_{(k,\sigma)\neq (k_0,\sigma_0)}
		\dyad{\gamma^\alpha_{k,\sigma}(t)}{\gamma^{\beta^*}_{k,\sigma}(t)}.
        \label{Eq: Field_state}
	\end{align}
    Equation (\ref{Eq: Field_state}) shows that the emitted field state is a weighted average of many coherent states driven with a SC driving field, all weighted by the distribution function $P(\alpha, \beta)$. We note that the right-hand side of Eq. (\ref{Eq: Field_state}) is completely specified, once the current element $\vec{j}_{i,i}^\xi(t)$ [and hence $\gamma_{k, \sigma}^\xi(t)$] is obtained and $P$ is specified. The phase factors with the time-dependent electronic energies will disappear under the choice of representation considered in Sec. \ref{Sec: Representation}.

\subsection{Representation of the driving fields} \label{Sec: Representation}

	We consider different types of driving fields. Specifically, we drive our model with coherent, Fock, thermal, and BSV light. As the generated photonic state [Eq. (\ref{Eq: Field_state})] was derived in a general representation, the choice of representation can be made freely which we will now specify for each of the driving fields. For Coherent and thermal light we use the Glauber-Sudarshan representation which takes $d\mu = d^2\alpha d^2\beta \delta^2(\alpha-\beta^*)$ \cite{Drummond1980} with corresponding distributions \cite{Gerry2004}
	\begin{align}
		P^{(\alpha_L)}_{\text{Coherent}}(\alpha) &= \delta^2(\alpha-\alpha_L), 		
		\label{Eq: P_coherent} 
		 \\
		P^{(\langle N \rangle)}_{\text{Thermal}}(\alpha) &= \frac{1}{\pi\mean{N}}\exp\!\left(-\frac{\abs{\alpha}^2}{\mean{N}}\right),
		\label{Eq: P_thermal}
	\end{align}
	where $\alpha_L$ is the coherent state amplitude of the driving field and $\langle N \rangle$ is the mean photon number for coherent and thermal light, respectively. 	
	The Glauber-Sudarshan representations for Fock and BSV light are highly singular \cite{Gerry2004, Walls2008} and therefore not computationally useful. Instead, we use the positive $P$ representation for these types of quantum light \cite{Gerry2004}. In the positive $P$ representation $d\mu = d^2 \alpha d^2 \beta$, with the corresponding positive $P$ distribution given by
	\begin{equation}
		P(\alpha, \beta) = (4\pi)^{-1}\exp\!(-\abs{\alpha-\beta^*}^2/4)Q[(\alpha+\beta^*)/2],
		\label{eq:Positive_P_general_in_Q}
	\end{equation}  
	with 
	    \begin{align}
		Q(\alpha) = \frac{1}{\pi}\mel{\alpha}{\hat{\rho}}{\alpha}
	\end{align} 
	the Husimi $Q$ function \cite{Kim1989}. Following Ref. \cite{Gorlach2023}, the interaction volume is assumed to be small and the average photon number of the driving field to be large, so the positive $P$ function in Eq. (\ref{eq:Positive_P_general_in_Q}) is approximated as 
	\begin{align}
		P(\alpha,\beta) \approx \delta(\alpha-\beta^*) Q\left(\frac{\alpha+\beta^*}{2}\right).
		\label{Eq: Positive_P_approx}
	\end{align}
	We shall refer to Eq. (\ref{Eq: Positive_P_approx}) as the approximative positive $P$ (APP) representation. Within the APP we thus describe Fock and BSV light as \cite{Kim1989}

	\begin{align}
		P^{(n)}_{\text{Fock}} (\alpha, \beta) \approx& ~  \delta(\alpha-\beta^*) P^{(n)}_{\text{Fock}}(\alpha ),
\end{align}
where
\begin{align}
		 P^{(n)}_{\text{Fock}} (\alpha) =  Q^{(n)}_{\text{Fock}} (\alpha)=& \frac{1}{\pi}e^{-\abs{\alpha}^2}\frac{\abs{\alpha}^{2n}}{n!}, 
		 \label{Eq: P_fock}
\end{align}
and similarly
\begin{align}
		P^{(r)}_{\text{BSV}} (\alpha, \beta) \approx& ~ \delta(\alpha-\beta^*) P^{(r)}_{\text{BSV}}(\alpha),		
\end{align}
where
\begin{align}
		 P^{(r)}_{\text{BSV}} (\alpha) &=  Q^{(r)}_{\text{BSV}} (\alpha) \nonumber \\
		 &= \frac{1}{\pi \cosh r}\exp\!\left(-\frac{2[\text{Re}(\alpha)]^2}{1+e^{-2r}}-\frac{2[\text{Im}(\alpha)]^2}{1+e^{2r}}\right)
     \label{Eq: P_bsv}
	\end{align}
for a Fock state $\ket{n}$ and for a squeezed vacuum state $\hat{S}(r)\ket{0}$, respectively, with $\hat{S}$ being the squeezing operator \cite{Gerry2004} and where $r$ is chosen to be real for simplicity. In both Eqs. (\ref{Eq: P_fock}) and (\ref{Eq: P_bsv}), we have defined the $P$ function within the APP representation to be the appropriate Husimi $Q$ function.

As the four representations of light in Eqs. (\ref{Eq: P_coherent}, \ref{Eq: P_thermal}, \ref{Eq: P_fock}, \ref{Eq: P_bsv}) along with the corresponding expression for $d\mu$ are diagonal in phase space [Eqs. (\ref{Eq: P_fock}) and (\ref{Eq: P_bsv}) only diagonal within the APP], we can rewrite the field state in Eq. (\ref{Eq: Field_state}) for the cases considered as
	\begin{align}
		\hat{\rho}_F(t) =
		\int d^2\alpha
		&P(\alpha)
		\dyad{\gamma^\alpha_{k_0,\sigma_0}(t)+\alpha}{\gamma^{\alpha}_{k_0,\sigma_0}(t)+\alpha}
		\notag
		\\
		& \bigotimes_{(k,\sigma)\neq (k_0,\sigma_0)}
		\dyad{\gamma^\alpha_{k,\sigma}(t)}{\gamma^{\alpha}_{k,\sigma}(t)},
		\label{Eq: P_field}
	\end{align}
	where the $P$ distribution, $P(\alpha)$, for the driving field can describe either of the four types of driving fields in Eqs. (\ref{Eq: P_coherent}, \ref{Eq: P_thermal}, \ref{Eq: P_fock}, \ref{Eq: P_bsv}). From Eq. (\ref{Eq: P_field}), expectation values of interest can be calculated. These expectation values are exact for coherent and thermal light driving fields, while they are approximate for Fock light and BSV driving fields, due to the APP representation. A discussion of the accuracy of the APP representation in relation to different observables is given in Sec. \ref{Sec: Accuracy_of_Q} and App. \ref{App: squeez_in_Q}.
	

\section{The harmonic spectrum}\label{Sec: harmonic_spectrum}
	Using Eq. (\ref{Eq: P_field}), the harmonic spectrum can be found by considering the energy in the photonic degrees of freedom $\mathcal{E} = \sum_{\vec{k},\sigma}\omega_k \mean{\hat{a}_{\vec{k},\sigma}^\dag \hat{a}_{\vec{k},\sigma}}$ as in Refs. \cite{Gorlach2020,Gorlach2023,Lange2024a}. Letting $\sum_{\vec{k}} \mapsto V/(2\pi c)^{3} \int d\omega \: \omega^2 \int d\Omega $, we can write out the sum over polarizations $\sigma$ and express the spectrum as the emitted energy per frequency. Furthermore, we disregard the driving field in the laser mode, i.e., we take  $\ket{\gamma^{\alpha}_{k_0,\sigma_0}(t) + \alpha} \rightarrow  \hat{D}^\dagger(\alpha)\ket{\gamma^{\alpha}_{k_0,\sigma_0}(t) + \alpha} = \ket{\gamma^{\alpha}_{k_0,\sigma_0}(t)}$ in Eq. (\ref{Eq: P_field}) since we are only interested in the generated field. The spectrum can then be expressed as
	\begin{align}
		S(\omega) \propto \int d^2\alpha P(\alpha) S_{SC}(\omega;\alpha)
		\label{Eq: spektrum}
	\end{align}
	with
	\begin{align}
	 S_{SC} (\omega;\alpha) = \left| \int_{-\infty}^\infty dt' j^\alpha_{i,i}(t') e^{i\omega t'} \right|^2 
	 \label{Eq: spectrum_semiclassical}
	\end{align}
	being the expression for the SC spectrum. Looking at Eq. (\ref{Eq: spektrum}), this leads to the same interpretation as in Ref. \cite{Gorlach2023}: The spectrum is a weighted average of SC spectra [Eq. (\ref{Eq: spectrum_semiclassical})] each driven with a classical field characterized by the coherent-state parameter $\alpha$ and weighted by a $P(\alpha)$ distribution.  
	
\subsection{The Floquet limit} \label{Sec: Floquet_limit}
    To analyze the harmonic spectrum [Eq. (\ref{Eq: spektrum})] further, we assume a time-periodic vector potential, i.e., a vector potential with no envelope. 
    The vector potential is written as $A^{\alpha}_{cl}(t)=2g_0 \omega_0^{-1/2}\abs{\alpha}\sin(\omega_0 t - \phi)$ where $\phi$ is the phase of $\alpha$. We assume that the conduction band is filled symmetrically around $q=0$ with an equal number of electrons with both spin orientations. Then, the dispersion function in Eq. (\ref{Eq: general_current}) can be expanded via the trigonometric addition identities, where the $\sin(alq)$ terms are discarded, since these are odd functions of $q$. From these assumptions the current matrix element $\vec{j}^{\alpha}_{i,i}(t)$ from Eq. (\ref{Eq: current_matrix_element}), with $\hat{\vec{j}}^{\alpha}(t)$ from Eq. (\ref{Eq: general_current}), becomes
	\begin{align}
		j^\alpha_{i,i}(t) = -2a \sum_l l b_l 
		\left[\sum_{q} \cos(alq)\right]
		\sin\!\left[ a l A^\alpha_{cl}(t)\right],
		\label{Eq: j_Floquet_analytical}
	\end{align}
	along the direction of the chain. In Eq. (\ref{Eq: j_Floquet_analytical}) the sum over crystal momentum states $q$ is now only over the occupied states. 
	\par
	Looking at the expression for the spectrum [Eqs. (\ref{Eq: spektrum}) and (\ref{Eq: spectrum_semiclassical})], we note that it is proportional to the square of Fourier transformed current. To proceed analytically, we use the Jacobi-Anger expansion and perform the Fourier transform of the time-dependent part of Eq. (\ref{Eq: j_Floquet_analytical}) as done in Ref. \cite{Lange2024}
	\begin{align}
		\int_{-\infty}^\infty &dt' \sin\!\left( al A^\alpha_{cl}(t')\right)e^{i\omega t'}
		=
		\sqrt{\pi}i \sum_{n=1}^\infty 
		J_{2n-1}(l\tilde{g}_0\abs{\alpha}) 
		\notag
		\\
		\times &e^{i\frac{\omega}{\omega_0}\phi}\left\{
		\delta \! \left[ \omega + (2n-1)  \omega_0 \right] -
		\delta \! \left[ \omega - (2n-1)   \omega_0 \right]
		\right\},
		\label{Eq: current}
	\end{align}
	where $J_n$ denotes the $n$'th order Bessel function of the first kind, $\delta$ denotes the delta function and $\tilde{g}_0 = 2ag_0/\sqrt{\omega_0}$ is a lattice-modified coupling constant. 
	Taking the square norm, the products of delta functions for different $n$'s must vanish. Hence, any phase on the delta functions vanishes. Furthermore, as we restrict ourselves to positive frequencies, the terms $\delta\left[\omega+(2n-1)\omega_0\right]$ may be discarded. Therefore, by inserting Eqs. (\ref{Eq: j_Floquet_analytical}) and (\ref{Eq: current}) into Eqs. (\ref{Eq: spektrum}) and (\ref{Eq: spectrum_semiclassical}), the generated HHG spectrum in the Floquet limit is given as
	\begin{align}
		S(\omega) \propto & \!\!\!\!\! \!\!\sum_{n=1, 3, 5, \dots}^\infty \!\!\!\!\!
		\omega^2\delta \left[ \omega - n   \omega_0 \right] \!\! \int \!\!d^2\alpha P(\alpha)\!
		\left[\sum_l C_l J_{n}(l\tilde{g}_0\abs{\alpha})\right]^2 \!,
		\label{Eq: spektrum_floquet}
	\end{align}
    where we collect crystal specific properties in the coefficient $C_l = l b_l \sum_q \cos(alq)$.
	In Eq. (\ref{Eq: spektrum_floquet}), the selection rule for the odd harmonics is clearly seen. We also see that the size of the $n$'th harmonic peak is determined by the integral of the product between the $P$ distribution function and a weighted sum of Bessel functions of order $n$. As we now show, analysis of these integrals can be used to gain insights into the characteristics of the spectrum.
	
\subsection{Simulated spectra}\label{Sec: simulate}
    \begin{figure}[t]
        \includegraphics[width=\linewidth]{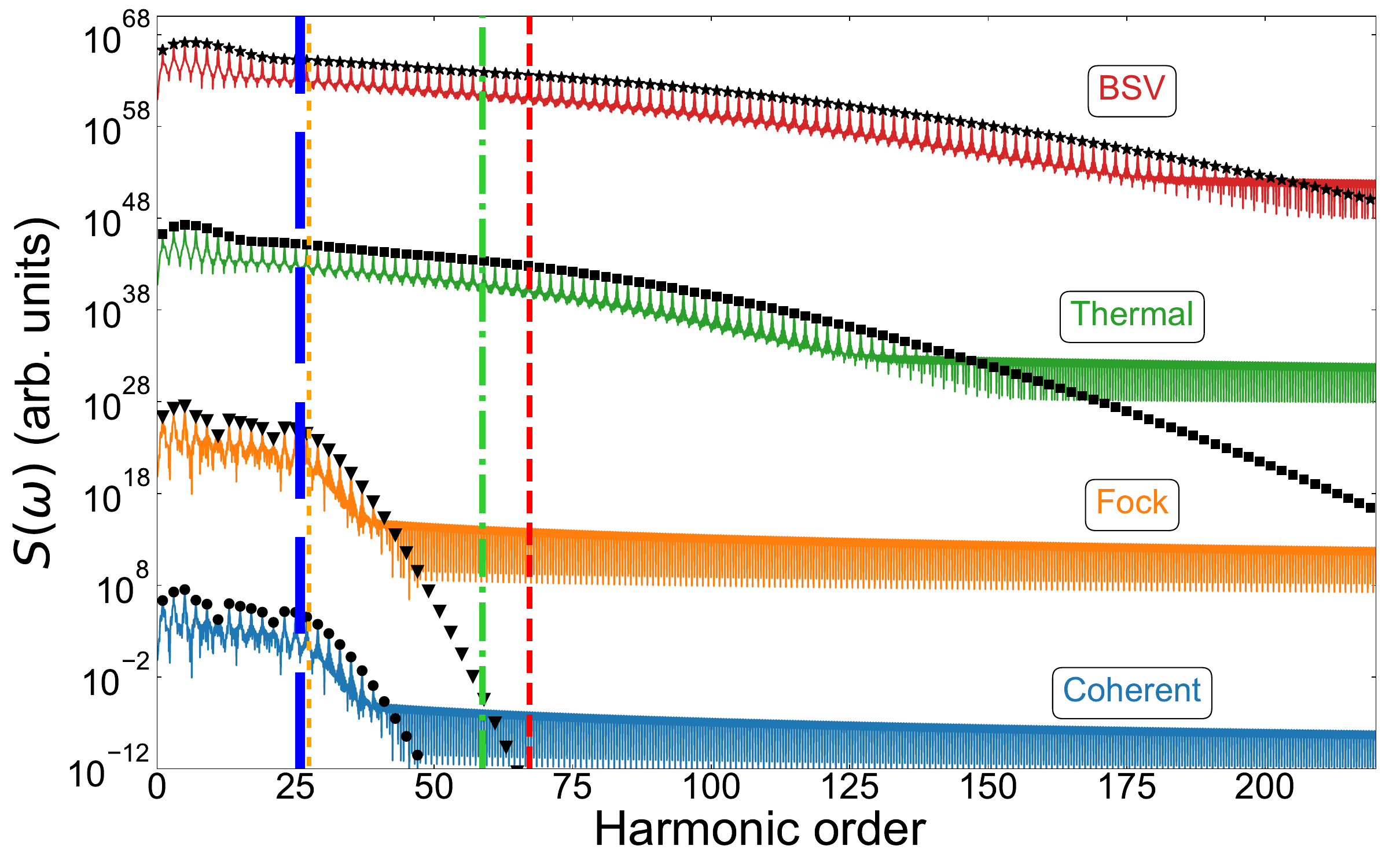}
        \caption{Simulated HHG spectra using the parameters specified in Sec. \ref{Sec: simulate}. Graphs [coherent (blue), Fock (orange), thermal (green), and BSV (red)] are simulated from Eq. (\ref{Eq: spektrum_polar}) with at $10$ cycle flat-top pulse with additional $3$ cycle turn-on and turn-off. For visual clarity, the spectrum for HHG driven by Fock light is multiplied by $10^{20}$, for Thermal light $10^{40}$, and for BSV $10^{60}$. Black markers [coherent (dots), Fock (triangles), thermal (squares), and BSV (stars)] are simulated in the Floquet limit using Eq. (\ref{Eq: spektrum_floquet}). Vertical lines [coherent (blue, long dashed), Fock (orange, dotted), thermal (green, dash-dotted), and BSV (red, short dashed)] denote the predicted cutoff calculated with Eq. (\ref{Eq: cutoff}).}
        \label{Fig: sim_spectrum_log}
    \end{figure}
    \par
	From Eqs. (\ref{Eq: spektrum}) and (\ref{Eq: spektrum_floquet}) the HHG spectra can be computed for a quantum state of driving light, given the appropriate $P$ functions. To simplify the computation, we note that for the HHG spectrum in the Floquet limit [Eq. (\ref{Eq: spektrum_floquet})] the only dependence on the phase of $\alpha$ in the integrand is contained in the $P$ function. Assuming an adiabatic evolution of the envelope of the driving field [which is already a necessary assumption to apply an envelope to Eq. (\ref{Eq: Classical_vector_potential})], we can assume that the norm square of the Fourier transform of the current matrix element is independent of the phase of $\alpha$. Therefore, we may write the harmonic spectrum of Eq. (\ref{Eq: spektrum}) using polar coordinates for integration as
    \begin{align}
        S(\omega) \propto 
        \omega^2 
        \int_0^\infty d\abs{\alpha}
        \left|
			\int_{-\infty}^\infty dt' j^\alpha_{i,i}(t') e^{i\omega t'}
		\right|^2 P(\abs{\alpha}),
        \label{Eq: spektrum_polar}
    \end{align} 
where we have defined the radial distribution as 
    \begin{align}  
P(\abs{\alpha}) =     \abs{\alpha} \int_0^{2\pi}d\phi P(\alpha),
\label{Eq: rad_dist}
     \end{align}
  and  where $\phi$ again denotes the phase of $\alpha$. As such, the angular integral can be performed without relation to frequency $\omega$, which lowers the dimensionality of computation. This approximation is valid if the width of the individual harmonic peaks overlap minimally, which is the case for the $\sim 16$ laser cycles pulse that we use for simulations. Using Eq. (\ref{Eq: spektrum_polar}), we calculate the spectra where the electrons are driven by a flat top pulse with $10$ cycles of constant amplitude with an additional $3$ cycle $\sin^2$ turn-on and turn-off as in Ref. \cite{Lange2024}. We use the $P$ functions for coherent [Eq. (\ref{Eq: P_coherent})], thermal [Eq. (\ref{Eq: P_thermal})], Fock [Eq. (\ref{Eq: P_fock})], and BSV [Eq. (\ref{Eq: P_bsv})] driving fields. 

    \par
    In the simulations, we use $g_0=4 \times 10^{-8}$ a.u. and $\omega_0   = 0.005$ a.u. as in Ref. \cite{Lange2024a}, a mean photon number of the driving fields $\mean{\hat{a}^\dagger_{\vec{k}_0, \sigma_0} \hat{a}_{\vec{k}_0, \sigma_0}} =7.35\times10^{11}$ (a coherent field with this photon number has the intensity $I = 8.26\times10^{11} $ $\text{W/cm}^2$). We use the parameters for the first conduction band in a ZnO crystal along the $\Gamma-M$ direction \cite{Vampa2015a} populated by $L=10$ electrons. Clearly, the characteristics of the results are independent of the number of electrons. In this model, the lattice constant is $a=5.32$ a.u. and the Fourier coefficients are $b_1 = -0.0814$ a.u., $b_2 = -0.0024$ a.u., $b_3 = -0.0048$ a.u., $b_4 = -0.0003$ a.u. and $b_5 = -0.0009$ a.u. The periodicity is chosen such that $\Delta q = \frac{2\pi}{10a}$ and $q_{max} = 2\Delta q$. For the BSV light, the squeezing parameter is obtained from the relation  $\mean{\hat{a}^\dagger_{\vec{k}_0, \sigma_0} \hat{a}_{\vec{k}_0, \sigma_0}} = \sinh^2(r)$, yielding $r = 14.35$ in our simulations. In Ref. \cite{Manceau2019} the production of BSV fields reaching $r = 15.3\pm 0.5$ was reported, and the BSV field considered in this work is therefore experimentally realizable. 
    \par
    The spectra in Fig. \ref{Fig: sim_spectrum_log} show the overall same behavior as the spectra in Ref. \cite{Gorlach2023} for atomic gasses. 
    The spectrum generated from a Fock-state driving field is visually identical to the coherently driven spectrum, both with a clear cutoff. On the other hand, the spectra for thermal and BSV driving fields are much broader with less clearly defined cutoffs but with the generation of much higher harmonics. Coherent and Fock fields produce the same field because the radial part of the $P$ function for coherent light is a delta function and similarly the radial part of the $P$ function for a Fock state is approximately a delta function for large photon numbers \cite{Gorlach2023}. This is further discussed in Secs. \ref{Sec: cutoff} and \ref{Sec: Accuracy_of_Q}.
    \par
    In Fig. \ref{Fig: sim_spectrum_log}, the finite pulse [Eq. (\ref{Eq: spektrum})] and Floquet limits [Eq. (\ref{Eq: spektrum_floquet})] are also compared, where the harmonic peaks obtained from the Floquet calculation are given by the black markers. The heights of the harmonic peaks are identical in the two calculations up to a global constant. Longer pulse durations just create sharper peaks. This is congruent with studies of inter- and intracycle aspects of HHG \cite{Andersen2024}.

\subsection{Cutoff}\label{Sec: cutoff}
	Figure \ref{Fig: sim_spectrum_log} shows that the characteristics of the spectra pertaining to the harmonic peaks can be interpreted in the Floquet limit. Via Eq. (\ref{Eq: spektrum_floquet}), the magnitude of the $n$'th harmonic is proportional to the integral of the product between the distribution function of the driving field, $P(\alpha)$ and the square sum of Bessel functions of order $n$. 
    The Bessel function $J_n(x)$ quickly vanishes for $x<n$ \cite{Andersen2024}. Hence, if the $P$ function is only nonzero within the region where the arguments of the Bessel function are smaller than its order, the overlap between these functions will vanish and there will be no harmonic peak. 
    \par 
   The argument of the Bessel function is $l\tilde{g}_0\abs{\alpha}$, which means that the $\abs{\alpha}$-interval where all the Bessel functions of order $n$ vanish is $\abs{\alpha} < n/(l_{max}\tilde{g}_0)$, where $l_{max}$ denotes the highest significant order in the Fourier expansion of the conduction band, which in the simulations of Fig. \ref{Fig: sim_spectrum_log} is $l_{max}=5$ (see Ref. \cite{Andersen2024} for a discussion of the relation between the Bessel functions and cutoff in intraband HHG). 
    \begin{figure}[t]
        \centering
        \includegraphics[width=\linewidth]{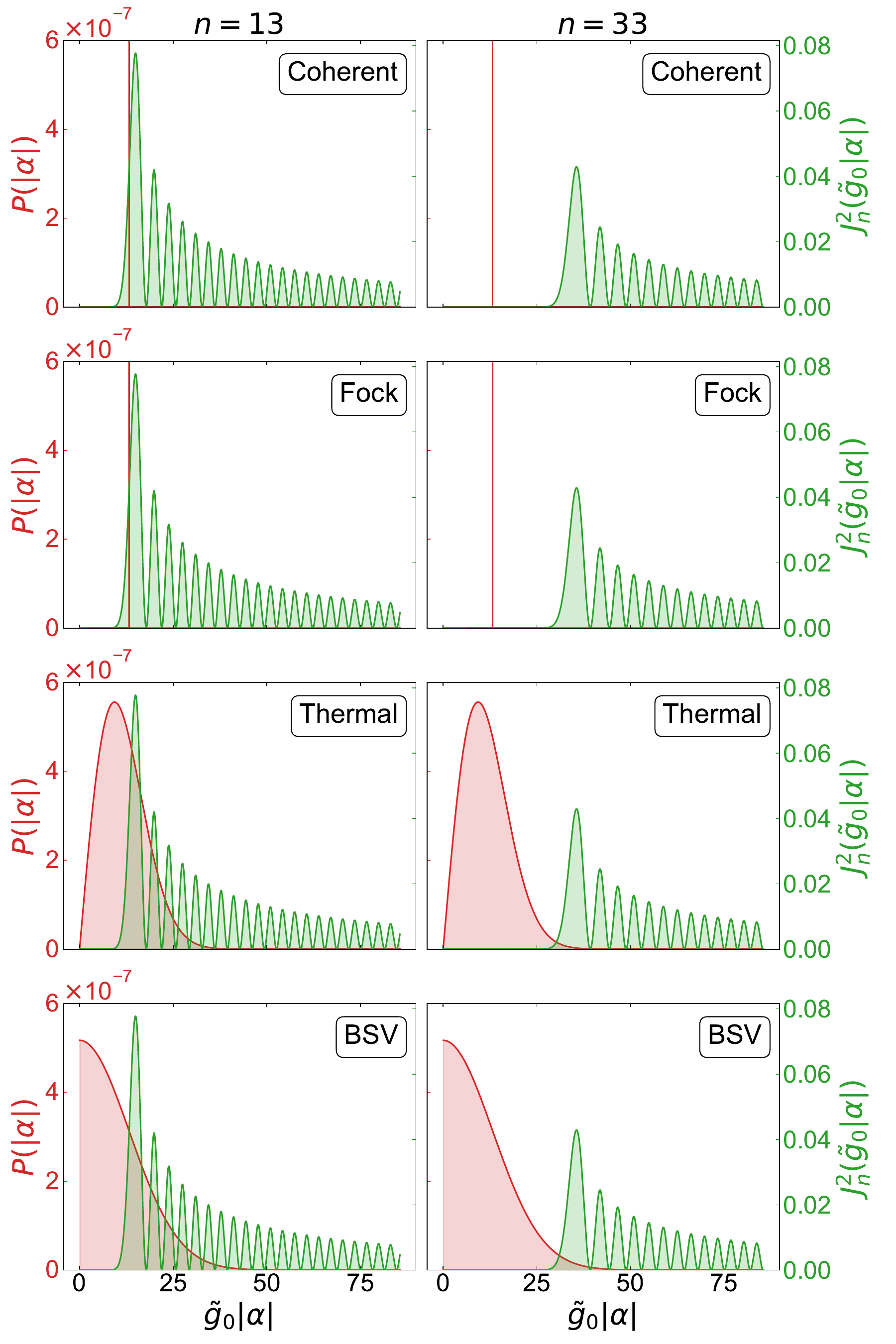}
        \caption{Illustration of the radial distribution functions, $ P(|\alpha|)$ of Eq. (\ref{Eq: rad_dist}) for a coherent, Fock, thermal, and BSV driving field at the same mean photon number as used in Fig. \ref{Fig: sim_spectrum_log} plotted against a Bessel function of order $13$ (left) and $33$ (right). The APP representation is used for the thermal and BSV fields. The behavior of $J_n(x)\approx 0$ for $x<n$ can be seen. For coherent and Fock states, there is no overlap for $n=33$. However, for the thermal field and even more so for the BSV field, a non-vanishing overlap is seen. Note that the $P(| \alpha |)$ distributions for the coherent and Fock states extend beyond the range of the ordinate.
        }
        \label{Fig: bessel_og_Q}
    \end{figure}
    The distributions for coherent and Fock states are narrow on the axis of $\abs{\alpha}$. In fact, the $P$ function for a coherent state is a delta function on $\abs{\alpha}$, and the APP function for a Fock state can be approximated as a delta function on $\abs{\alpha}$ \cite{Gorlach2023}. Therefore, for ascending harmonic orders, the regime of no overlap is quickly reached for coherent and Fock driving fields in comparison with thermal and BSV driving fields. This is depicted in Fig. \ref{Fig: bessel_og_Q}, where the square of the Bessel function $J_n^2(\tilde{g}_0 \abs{\alpha})$ and the $P$ function is plotted as a function of the argument $\abs{\alpha} \tilde{g}_0$ for harmonics $n= 11$ and $33$. 

   The quickly vanishing overlap for Fock and coherent driving fields explains why these two spectra are very similar in Fig. \ref{Fig: sim_spectrum_log}. The heights of the peaks and also the cutoff are determined only from the radial distributions $P(\abs{\alpha})$ [see Eq. (\ref{Eq: rad_dist})], which for coherent and Fock states are approximately equal. Furthermore, for coherent driving fields this cutoff is the same as found in Ref. \cite{Andersen2024}.
    For thermal and BSV fields, the distributions are broad and extend much further than their mean value (see Fig. \ref{Fig: bessel_og_Q}). Hence, even though the mean photon number might be the same, the broad distributions of these reach far larger values of $\abs{\alpha}$ and the overlap between $P(|\alpha|)$ and the Bessel functions is nonvanishing for higher harmonics yielding a higher cutoff. The gradual decrease of the tails also explains why the cutoff is less clearly defined (see Fig. \ref{Fig: sim_spectrum_log}).
    \par 
    If we were to define the cutoff ($\gamma_{\text{cutoff}}$) in the units of harmonic order for such a state, it makes sense from these considerations to write it as 
    \begin{align}
        \gamma_{\text{cutoff}} = l_{max}\tilde{g}_0(\mu_P + 3\sigma_P),
        \label{Eq: cutoff}
    \end{align}
    where $\mu_P$ and $\sigma_P$ are the mean and standard deviation, respectively, of the variable $\abs{\alpha}$ given the distribution $P(\alpha)$, which are equivalent to the mean and standard deviation of the radial distribution $P(\abs{\alpha})$ [see Eqs. (\ref{Eq: mu_P}) and (\ref{Eq: sigma_P}) for explicit expressions]. These quantities are therefore unitless and entirely determined by the choice of $P$. 
    The cutoffs are calculated numerically and plotted alongside the harmonic spectra in Fig. \ref{Fig: sim_spectrum_log} as vertical lines. This choice of cutoff is sensible, as most of the $P$ function will be contained within the domain of $\mu_P - 3\sigma_P \leq \abs{\alpha} \leq \mu_P + 3\sigma_P$, and it yields the correct cutoff for the coherent and Fock driving fields as seen in Fig. \ref{Fig: sim_spectrum_log}. However, as is also apparent from Fig. \ref{Fig: sim_spectrum_log}, the cutoff is not clearly defined for the thermal and BSV fields due to the more gradually vanishing overlap.

\subsection{Power Scaling Regime}\label{Sec: Power_scaling}
    In relation to experimental results like those presented in, e.g., Ref. \cite{Rasputnyi2024},
    we can, from Eq. (\ref{Eq: spektrum_floquet}), determine that the signal in the $n$'th odd harmonic is
    \begin{align}
        S (n\omega_0) \propto\int d^2\alpha P(\alpha)
		\left[\sum_l C_l J_{n}(l\tilde{g}_0\abs{\alpha})\right]^2.
		\label{Eq: harmonic_floquet}
    \end{align}
    Taylor expanding the Bessel functions to the lowest non-vanishing order gives $J_n(x) \approx x^n/(n!2^n)$. From this, it follows that
    \begin{align}
        &\left[\sum_l C_l J_{n}(l\tilde{g}_0\abs{\alpha})\right]^2
        \approx
        \frac{\tilde{g}_0^{2n}K_n}{(n!)^2 2^{2n}}\abs{\alpha}^{2n},
        \label{Eq: approx_bessel}
    \end{align}
    where $K_n = \sum_{l_1,l_2}l_1^{n}l_2^{n} C_{l_1}C_{l_2}$ is a geometric constant of the material, which is determined by the dispersion relation and the filling of the conduction band. Equation (\ref{Eq: approx_bessel}) is substituted into Eq. (\ref{Eq: harmonic_floquet}) to give (to lowest order)
    \begin{align}
        S(n\omega_0) \:\propto\:& K_n \frac{\tilde{g}_0^{2n}}{(n!)^2 2^{2n}} \int d^2\alpha P(\alpha) \abs{\alpha}^{2n}
        \notag
        \\
        =&
        K_n \frac{\tilde{g}_0^{2n}}{(n!)^2 2^{2n}} \mean{: \big(\hat{a}^\dagger_{\vec{k}_0, \sigma_0} \hat{a}_{\vec{k}_0, \sigma_0}\big)^n:}
        \notag
        \\
        =&
        K_n \frac{\tilde{g}_0^{2n}}{(n!)^2 2^{2n}} g^{(n)}(0) \mean{\hat{a}^\dagger_{\vec{k}_0, \sigma_0} \hat{a}_{\vec{k}_0, \sigma_0}}^n,
        \label{Eq: Power_scaling}
    \end{align}
    where $: \hat{O} :$ denotes the normal ordering of the operator $\hat{O}$ and $g^{(n)}(0)$ is the $n$'th order single-mode normalized correlation function. Thus, we have recovered the power scaling law $\mean{\hat{a}^\dagger_{n\vec{k}_0, \sigma_0} \hat{a}_{n\vec{k}_0, \sigma_0}}\propto  \mean{ \hat{a}^\dagger_{\vec{k}_0, \sigma_0} \hat{a}_{\vec{k}_0, \sigma_0}}^n$ of the perturbative regime \cite{Rasputnyi2024}. Interestingly, we note that in this particular model, the $n$'th order correlation function of the driving field, $g^{(n)}(0)$, enters as a proportionality factor and dictates the height of the signal in the $n$'th harmonic. This is another way of showing how the quantum optical nature of the driving field affects the generated HHG spectrum. In other words, this means, that if one could precisely control the mean photon number of the driving field, the height of the $n$'th harmonic is determined by the quantum nature of the driving field via $g^{(n)}(0)$. We note for clarity that this scaling law only holds when the lowest-order approximation of the Bessel function is a valid approximation, see App. \ref{App: Bessel_approx}. 
    \begin{figure}[t]
		\includegraphics[width=\linewidth]{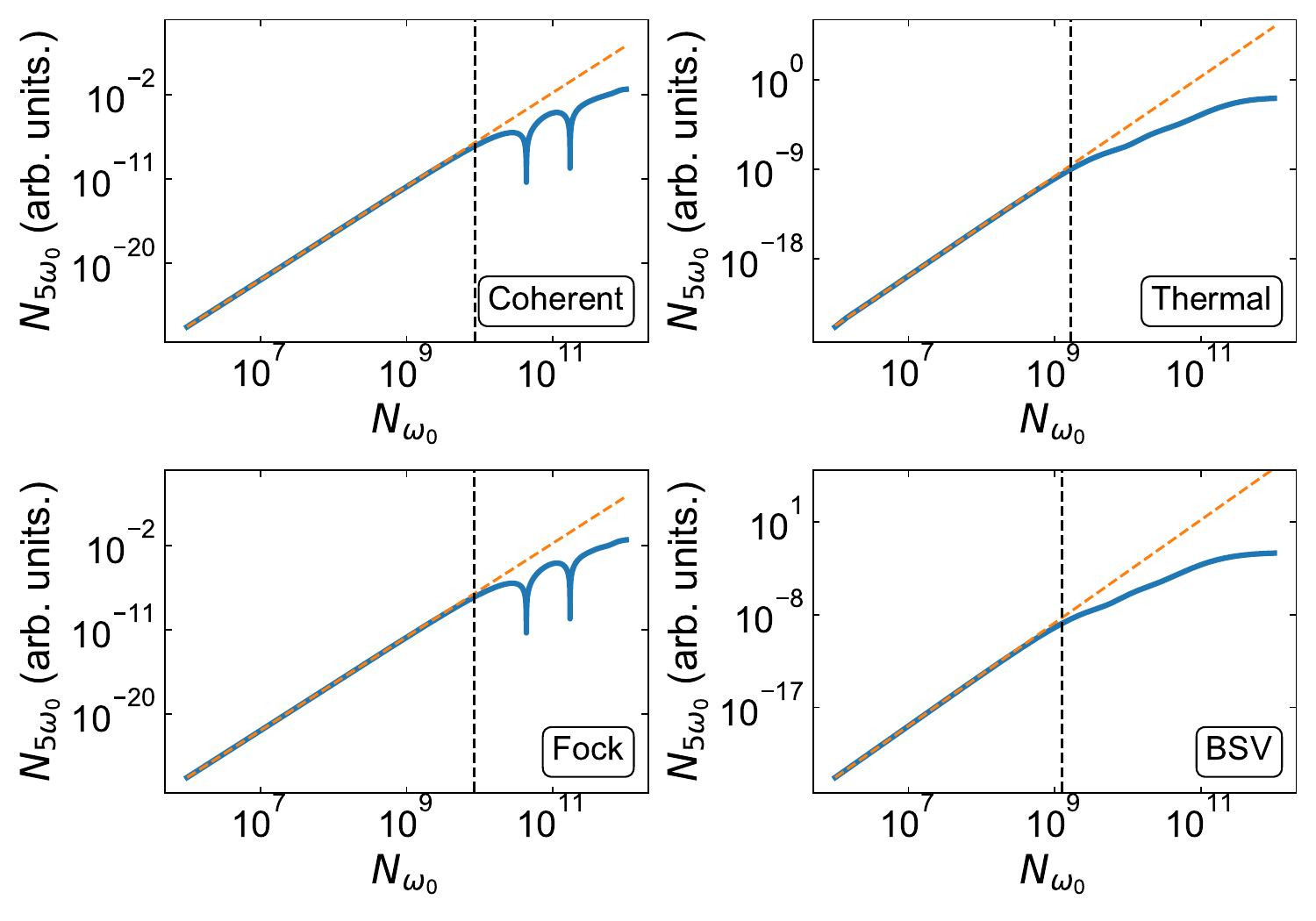}
		\caption{Solid, blue curves show the mean photon count \big($N_{n\omega_0}=\mean{\hat{a}^\dag_{n\vec{k}_0,\sigma_0}\hat{a}_{n\vec{k}_0,\sigma_0}}$\big) in the fifth harmonic as a function of the mean photon count of the driving field as computed from Eq. (\ref{Eq: harmonic_floquet}) using the $P$ representation of the different driving fields. The dashed, yellow line shows the perturbative power-scaling approximation computed from Eq. (\ref{Eq: Power_scaling}), and the black-dashed vertical line denotes the predicted range of applicability of the perturbative treatment.}
        \label{Fig: Powerscaling}
	\end{figure}
    \par
   In Fig. \ref{Fig: Powerscaling}, the peak intensity of the fifth harmonic is plotted as a function of the mean photon count in the driving field using Eq. (\ref{Eq: harmonic_floquet}) for the non-perturbative result and using Eq. (\ref{Eq: Power_scaling}) for the power-scaling approximation.  To determine when the lowest-order approximation of $J_m(x)$ is valid, we apply the Taylor remainder theorem, see details in App. \ref{App: Bessel_approx}. At the photon count, where the perturbative treatment is expected to break down based on the analysis in App. \ref{App: Bessel_approx}, we mark the cutoff for this expected range of applicability with a vertical dotted line in Fig. \ref{Fig: Powerscaling}. 
    
    \par

    As clearly seen in Fig. \ref{Fig: Powerscaling}, this analysis yields a good estimation of the range of applicability of a perturbative treatment. It is clear that this range is smaller for BSV and thermal driving fields. Our predictions can explain this behavior: since BSV and thermal states have much broader $P$ distributions, their standard deviation will be large. Therefore, these distributions will reach into the range where the lowest-order Taylor expansion is not valid at a lower mean photon count than for states with narrow distributions such as coherent and Fock states. In other words, due to the broad $P$ distributions of thermal and BSV states, these driving fields probe nonperturbative processes at lower intensities than coherent and Fock driving fields. 
    \par
    These findings are not immediately in line with Ref. \cite{Rasputnyi2024}, in which it is reported that HHG driven by BSV has a broader intensity range of the perturbative regime than HHG driven by coherent light. However, multiple factors make the comparison difficult. Firstly, only one of the harmonics reported in Ref. \cite{Rasputnyi2024} is an odd harmonic below the band gap and can be compared directly with our studies. For this harmonic, the threshold of optical damage for the coherent driving field is around the intensity cutoff of the perturbative regime for the BSV driving field. Below the optical damage threshold for the coherent driving field, a clear deviation from the power-scaling law is not apparent. That coherent driving should become nonperturbative at lower intensities is reported instead for the fourth harmonic, which does not appear in this work, as we consider spatially symmetric crystals. Finally, the results presented in Ref. \cite{Rasputnyi2024} may be affected by electrons being promoted across the bandgap even for the harmonics below the bandgap which is an effect not taken into account in this work.


\section{Time-resolved electric field}\label{Sec: E-field}
As is apparent in Eq. (\ref{Eq: P_field}), the derived density operator of the fields is time dependent. 
This allows for the evaluation of time-dependent observables such as the electric field, which bears relevance in relation to experimental work \cite{Sederberg2020, Zimin2022, Husakou2024, Bionta2021,Keathley2023}, where the temporal characteristics of the generated field may be observed. We therefore determine the electric field in the time domain within the considered model.
\par
The quantized electric field operator in the dipole approximation is given as
\begin{align}
	\hat{\vec{E}}(t) = i\sum_{\vec{k},\sigma}g_0\sqrt{\omega_{\vec{k}}}
	\left(
	\vec{e}_\sigma\hat{a}_{\vec{k},\sigma}e^{-i\omega_{\vec{k}}t}-h.c.
	\right)
	\label{Eq: E_felt_definition}
\end{align}
As the APP for Fock and BSV fields is an approximate description of the density matrix [Eq. (\ref{Eq: P_field})], we verify in App. \ref{App: E_field} that this approximation produces the correct electrical field for the different types of driving fields considered. 

\begin{figure}[t]
	\includegraphics[width=\linewidth]{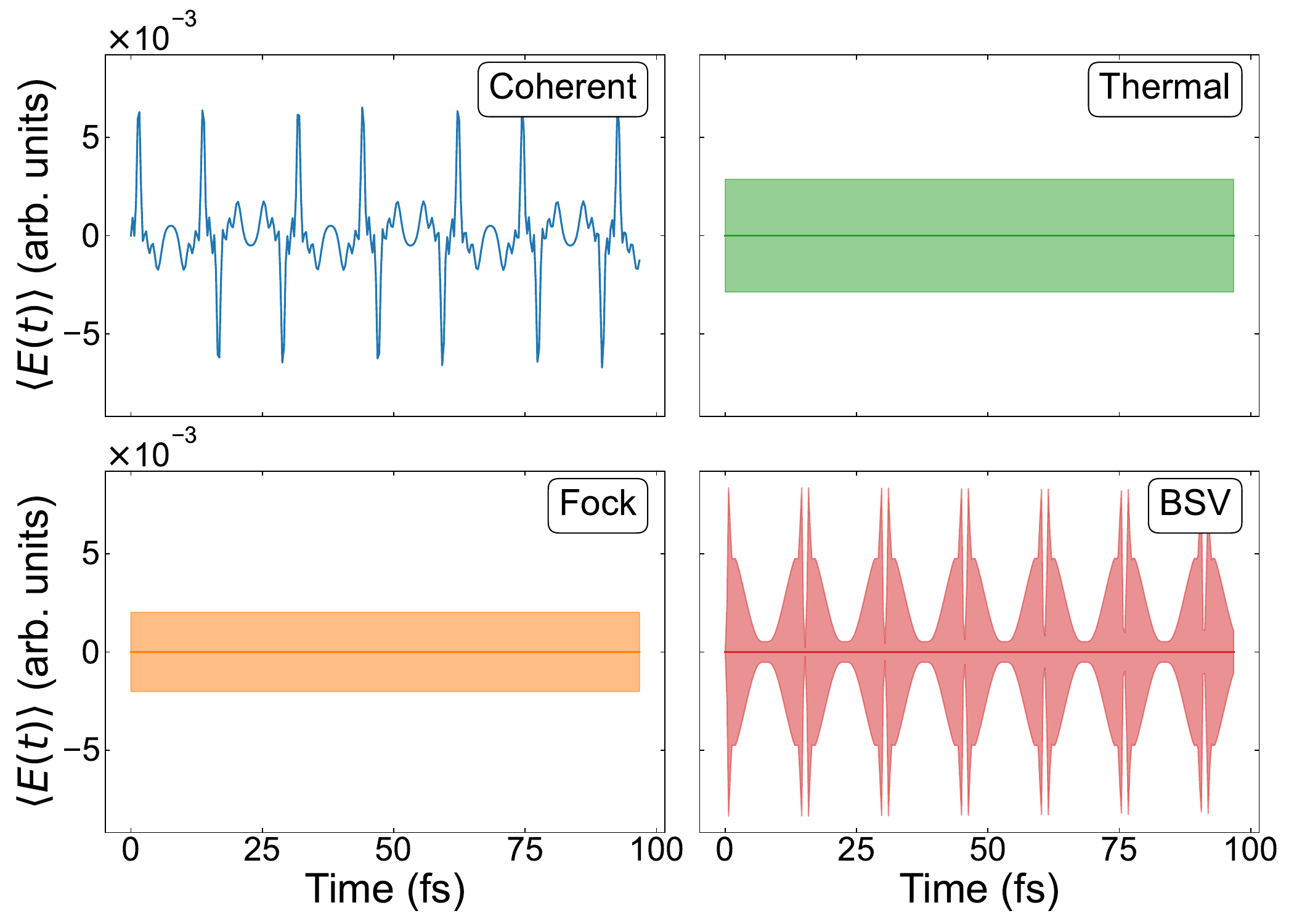}
	\caption{Time-resolved electric field generated by HHG for coherent, Fock, thermal, and BSV driving fields. Solid lines represent the expectation value of the electric field and the shaded area represents the uncertainty in the electric field neglecting the zero-point vacuum fluctuations. Simulated from Eqs. (\ref{Eq: tids_E_felt}) and (\ref{Eq: tids_E_felt_varians}). The same parameters are used as in Sec. \ref{Sec: simulate}.}
	\label{Fig: E_field_time_varians}
\end{figure}
We now investigate the generated electric field. Using that the coherent states are eigenstates of the annihilation operator, we use Eqs. (\ref{Eq: P_field}) and (\ref{Eq: E_felt_definition}) to determine the expectation value of the generated electric field as
\begin{align}
	&\mean{\hat{\vec{E}}(t)}_{\text{HHG}} = \Tr\!\left[\hat{\rho}(t)\hat{\vec{E}}(t)\right] 
	\notag \\
	= - \sum_{k,\sigma}g_0&\sqrt{\omega_k}
	\int d^2\alpha P(\alpha)  \:2\:\text{Im}\!\left[\hat{\vec{e}}_{k\sigma}\gamma^\alpha_{k\sigma}(t)e^{-i\omega_k t}\right],
\end{align}
where we, as in connection with Eq. (\ref{Eq: spektrum}), have neglected the $+\alpha$ from the laser in the driving mode as we are only interested in the generated field.
\par
Inserting Eq. (\ref{Eq: gamma}) and executing the sum over the polarizations and the solid angle integral, it is found that
\begin{align}
	\mean{\hat{\vec{E}}(t)}_{\text{HHG}}& = 
	-\frac{2Vg_0}{3\pi^2 c^3 } 
	\int d^2\alpha P(\alpha) 
	\notag\\
	\times\text{Im}&\! \left[
	-i
	\int^t_{t_i} dt' \vec{j}^\alpha_{i,i}(t')
	\int_0^\infty d\omega \: \omega^2 e^{i\omega(t'-t)}
	\right],
\end{align}
where the innermost integral can be recognized as the distribution $-\pi \frac{d^2}{dt^2}\delta(t'-t)$, when the imaginary part is taken. As $t\in[t_i,t]$ and by writing out $g_0$, we obtain
\begin{align}
	\mean{\hat{\vec{E}}(t)}_{\text{HHG}}& = 
	-\frac{4}{3c^3} 
	\int d^2\alpha P(\alpha) 
	\frac{d^2}{dt^2} \vec{j}^\alpha_{i,i}(t).
	\label{Eq: tids_E_felt}
\end{align}
Likewise, the variance of the generated field can be derived. Taking $\vec{j}_{i,i}$ along the one-dimensional chain of the model to be in the $z$-direction, we compute $\mean{E_z^2}$. Generally, this will yield four terms: The square of the generated fields, the square of the driving field, an interference term between the driving and generated field, and the zero-point fluctuations. As the generated field at frequencies different from $\omega_0$ is of interest, the other terms are neglected. For details, see App. \ref{App: E_field}. We then find that
\begin{align}
	\mean{\hat{E}_z^2}_{\text{HHG}} = \frac{16}{9c^6} \int d^2\alpha P(\alpha) \left[\frac{d^2}{dt^2}j_{ii}^\alpha(t)\right]^2,
\end{align}
which is used to determine the variance
\begin{align}
	\mean{\Delta \hat{E}_z^2}_{\text{HHG}} = 
	\frac{16}{9c^6} \left\{
	\int d^2\alpha P(\alpha) \left[\frac{d^2}{dt^2}j_{ii}^\alpha(t)\right]^2
	\right.
	\notag
	\\
	\left.
	-
	\left[\int d^2\alpha P(\alpha)\frac{d^2}{dt^2}j_{ii}^\alpha(t)\right]^2
	\right\}.
	\label{Eq: tids_E_felt_varians}
\end{align}
Note that the distribution function of the driving field, $P(\alpha)$, also enters into the expressions of the characteristics of the generated fields in Eqs. (\ref{Eq: tids_E_felt}) and (\ref{Eq: tids_E_felt_varians}), indicating that the quantum nature of the driving field is mapped onto the generated field. This mapping of field characteristics is in accordance with the simulations of the generated electric fields shown in Fig. \ref{Fig: E_field_time_varians} where both the electric field, $\mean{\hat{\vec{E}}(t)}_{\text{HHG}}$ [Eq. (\ref{Eq: tids_E_felt})], along with its uncertainty, $\sqrt{\mean{\Delta \hat{E}_z^2}_{\text{HHG}}}$ [Eq. (\ref{Eq: tids_E_felt_varians})], for different driving fields is shown. In Fig. \ref{Fig: E_field_time_varians}, it is apparent that the field generated from coherent light has the characteristics of coherent light. This is in accordance with the result of Ref. \cite{Lange2024a}, in which it is found that the harmonics of HHG are coherent when the driving field is coherent and the electrons of the medium are uncorrelated. For the BSV driving field, the generated field also has the characteristics of BSV fields, where the expectation value of the electric field is vanishing but it still has a large time-dependent uncertainty. Likewise, the generated fields from the Fock and thermal driving fields have vanishing expectations values of the fields with constant uncertainty, which are the same characteristics as their corresponding driving fields. 
\par
Estimating the width of the peaks in Fig. \ref{Fig: E_field_time_varians}, they are all at least $0.8$ fs corresponding to a maximum frequency of around $1$ PHz. This spectral resolution is within reach for experimental settings using, e.g., TIPTOE \cite{Bionta2021, Keathley2023}. As such, it should be feasible to measure such time-dependent fields, from which we can uncover information on how the quantum state of the driving field maps onto the quantum state of the emitted field via the process of HHG.


\section{Accuracy of the Approximative Positive $P$ Representation}\label{Sec: Accuracy_of_Q}
    Figures \ref{Fig: sim_spectrum_log} and \ref{Fig: Powerscaling} were computed using the APP representation for BSV and Fock light. The accuracy of these results therefore depends on the accuracy of the APP representation as introduced in Sec. \ref{Sec: Representation}. To illustrate why this is an important point, consider a Fock state $\ket{n}$, which is described by the density operator $\hat{\rho} = \int d^2\alpha P(\alpha)\dyad{\alpha}{\alpha}$ in the Glauber-Sudarshan representation where the $P$ function is \cite{Gerry2004}
    \begin{align}
        P^{(n)}_{\text{Fock, GS}}(\alpha) = \frac{e^{\abs{\alpha}^2}}{n!}\frac{\partial^{2n}}{\partial\alpha^{n}\partial\alpha^{*n}} \delta^{(2)} (\alpha).
        \label{Eq: fock_P}
    \end{align}
    Here, derivatives of the delta function are understood in the distributional sense. On the other hand, the APP representation for the Fock state is given by Eq. (\ref{Eq: P_fock}).
    From Eqs. (\ref{Eq: P_fock}) and (\ref{Eq: fock_P}) it is not trivial that $\hat{\rho} = \int d^2\alpha P^{(n)}_{\text{GS}}(\alpha)\dyad{\alpha}{\alpha} \approx \int d^2\alpha Q^{(n)}_{\text{Fock}}(\alpha)\dyad{\alpha}{\alpha}$ for large $n$. Indeed, it would be difficult to make the arguments of Secs. \ref{Sec: cutoff} and \ref{Sec: Power_scaling} using the highly non-singular $P$ function of Eq. (\ref{Eq: fock_P}), as one cannot make statements on the value of $P$ at a given $\alpha$. Hence, it is important to study if the APP is accurate for the observables that we consider and verify the APP on a case-by-case basis.
    
    In order to verify that the APP representation indeed yields reliable results (for BSV and Fock light), we test it on the incoming driving field. We find that it yields the correct photon number $\langle \hat{a}_{\vec{k}, \sigma}^\dagger \hat{a}_{\vec{k}, \sigma} \rangle$ in the limit of large photon number $n$ for both BSV and Fock light. Further, it reproduces the correct mean and variance of the time-dependent electric field as discussed in App. \ref{App: E_field}. However, the APP representation does not reproduce the correct  photon statistics or degree of squeezing for the driving field. This is due to the fact that the APP representation neglects the coherence of $\alpha$ and $\beta$ in the positive $P$ distribution [Eq. (\ref{eq:Positive_P_general_in_Q}), see also Eq. (\ref{Eq: Positive_P_approx})]. This shows a limitation of the APP and we do therefore not consider the photon statistics or squeezing of the emitted field. Further calculations and discussion on the validity of the APP representation is given in App. \ref{App: squeez_in_Q}.                                   


\section{Conclusion}\label{Sec: conclusion}
In this work, we presented a derivation of the quantum optical state of light emitted from intraband HHG driven by quantum light. Using this quantum optical state for the emitted field, we calculated the harmonic spectrum for different types of driving fields: coherent, Fock, thermal, and BSV light. These results were compared to the Floquet limit which produced identical spectra. Using the analytical Floquet limit, we studied the harmonic spectrum and predicted the harmonic cutoff for a given type of driving field. In this limit, the spectrum is expressed analytically in terms of Bessel functions $J_n(l \tilde{g}_0 \lvert \alpha \rvert)$, with arguments that depend on the dispersion relation of the generating crystal as well as the strength of the driving field. We found that the harmonic cutoff is dictated by the overlap between these Bessel functions and the $P$ distributions used to describe the driving field. This enabled us to explain why the cutoff is less clearly defined when driving with a thermal or BSV field than for a coherent or Fock field. Furthermore, we estimated the intensity range of the driving field for which the harmonic peaks can be treated perturbatively. From this analysis, we found that fields with broad coherent phase-space distributions can probe nonperturbative processes with a lower intensity than fields with narrow distributions such as coherent and Fock fields.

Utilizing that the derived field state is time dependent, we also derived expressions for the expectation value of the generated electric field and its variance. In the single-band model of a solid considered in this paper, we found that the temporal characteristics of the driving field are mapped to the generated field. Since these generated time-dependent fields vary on a timescale that can be observed experimentally, this presents potential for experimental studies on how the HHG process maps the quantum state of the driving field onto the generated field, as temporal characteristics depend on the quantum state of the field.  

We also discussed the accuracy of approximating the nondiagonal positive $P(\alpha, \beta)$ representation of a quantum state by an approximate positive $P$ description for the Fock and BSV driving fields. We confirmed that this approximation yields the correct harmonic spectra (in the limit of large photon numbers) and time-dependent electric fields for the driving fields before calculating these observables for the emitted field. However, this approximation does not yield the correct photon statistics or degree of squeezing for the driving field and we did therefore not consider these observables for the emitted field. Going beyond the APP and using the full positive $P(\alpha, \beta)$ is to be considered in future work, such that the nonclassical observables of the emitted light can be investigated in more detail.

The model of solids considered in this work is a simple one-dimensional single-band model. It is therefore relevant for future work to extend the description of the electronic system to include, e.g., multiple bands, topological effects and electron correlation. However, in such extensions it would not generally be possible to make the same analytical considerations regarding the induced current as in this work without further approximations. More specifically, the current operator would not generally commute with the Hamiltonian as it does in the model of this work and hence all cross current matrix elements would need to be considered which greatly increases the complexity of the problem. Approaches to get around this could be considered, such as using a Markov state approximation (see Ref. \cite{Lange2024b}), which can accurately capture features such as the generated spectrum, including squeezing and photon statistics.


\appendix

\section{Observables using the approximative positive $P$ representation}\label{App: squeez_in_Q}

In this appendix, we discuss which observables of the driving field are reproduced by using the APP representation for the Fock and BSV driving fields in Eqs. (\ref{Eq: P_fock}) and (\ref{Eq: P_bsv}).
The approximative APP for the density matrix for the driving field is then 
\begin{align}
	\hat{\rho}^{\text{(APP)}} = \int d^2\alpha~ Q(\alpha) \dyad{\alpha}{\alpha},
\end{align}
where $Q(\alpha)$ represents either Fock or BSV driving fields.

First, we calculate the mean photon number. The spectrum for a Fock state is determined by
\begin{align}
	\mean{\hat{a}^\dagger \hat{a}}^{(\text{APP})}_{\text{Fock}} &=  \Tr\big[\hat{a}^\dagger \hat{a} \hat{\rho}^{(\text{APP})} \big] \nonumber \\
	&=\int d^2\alpha~ Q^{(n)}_{\text{Fock}}(\alpha)  \abs{\alpha}^2 \nonumber \\
	&= \int_0^{2\pi} d\phi \int_0^\infty d\abs{\alpha} ~\frac{1}{\pi}e^{-\abs{\alpha}^2}\frac{\abs{\alpha}^{2n}}{n!} \abs{\alpha}^2, \nonumber \\
	&= n \nonumber \\
	&= \mean{\hat{a}^\dagger \hat{a}}^{\text{Exact}}_{\text{Fock}} 
\end{align}
which shows that the APP gives the correct result for the spectrum. Similarly, we find for a BSV state
\begin{align}
	\mean{\hat{a}^\dagger \hat{a}}^{(\text{APP})}_{\text{BSV}} &=  \Tr \big[\hat{a}^\dagger \hat{a} \hat{\rho}^{(\text{APP})} \big] \nonumber \\
	&= \int_{-\infty}^{\infty} dx \int_{-\infty}^\infty dy ~ Q^{(r )}_{\text{BSV}} (x^2 + y^2) \nonumber \\
	&= \cosh^2(r) \nonumber \\
	&\neq \mean{\hat{a}^\dagger \hat{a}}^{\text{Exact}}_{\text{BSV}},
	\label{Eq: a_dagger_a_BSV_Q_rep}
\end{align}
where we have used $\alpha = x + iy$. Equation (\ref{Eq: a_dagger_a_BSV_Q_rep}) differs from the exact result for a squeezed vacuum state which is $\mean{\hat{a}^\dagger \hat{a}}^{\text{Exact}}_{\text{BSV}} = \sinh^2(r)$ \cite{Gerry2004}. However, in the limit of a large photon number, we see that $\sinh^2(r) =   \cosh^2(r) -1 \approx \cosh^2(r)$ approximately yielding the correct result. Thus, as the spectrum is proportional to the mean photon number, we conclude that the spectra for the driving fields are correctly reproduced by the APP representation and we extend the analysis to the emitted HHG spectrum with this representation in Sec. \ref{Sec: harmonic_spectrum} in the main text. 

We now consider the photon statistics for the Fock driving field. Here we consider the Mandel-$\mathcal{Q}$ parameter (not to be confused with the Husimi-Q function) given by

\begin{align}
	\mathcal{Q} &= \dfrac{\mean{\hat{n}^2} - \mean{n}^2}{\mean{n}} - 1 \nonumber \\
	&= \dfrac{\mean{\hat{a}^\dagger \hat{a}^\dagger \hat{a} \hat{a}} - \mean{\hat{a}^\dagger \hat{a}}^2}{\mean{\hat{a}^\dagger \hat{a}}}.
\end{align}
Calculating the second moment in the APP representation, we find

\begin{align}
	\mean{\hat{a}^\dagger \hat{a}^\dagger \hat{a} \hat{a}}_{\text{Fock}}^{(\text{APP})} = n^2 + 3n + 2,
\end{align}
yielding a Mandel-$\mathcal{Q}$ parameter of $\mathcal{Q}^{(\text{APP})}_{\text{Fock}} = 3 + 2/n$, which does not match the exact result of $\mathcal{Q}^{\text{Exact}}_{\text{Fock}} = -1$. This is a significant error, as the APP result predicts super-Poissonian statistics for a state with sub-Poissonian statistics.

Similarly, we consider the degree of squeezing for the BSV driving field. This is done through minimizing the variance of the quadrature operator $\hat{X}(\theta) = \frac{1}{2}\left(\hat{a}e^{-i\theta} + \hat{a}^\dag e^{i\theta}\right) $ for $\theta\in [0,\pi)$. The variance of this operator is 
\begin{align}
	\mean{\Delta\hat{X}^2(\theta)} = \frac{1}{4}&\left[
	e^{-2i\theta}\left( \mean{\hat{a}^2} -\mean{\hat{a}}^2 \right)
	+
	e^{2i\theta}\left( \mean{\hat{a}^{\dag2}} -\mean{\hat{a}^\dag}^2 \right) \right.
	\notag
	\\
	&\left. +
	2\left(
	\mean{\hat{a}^\dag\hat{a}}- \mean{\hat{a}}\mean{\hat{a}^\dag}
	\right)
	+ 1
	\right],
	\label{variance_general}
\end{align}
Calculating the moments of the photonic operators, we find that in
\begin{align}
	\mean{\hat{a}}_{\text{BSV}}^{(\text{APP})} &= \mean{\hat{a^\dagger}}_{\text{BSV}}^{(\text{APP})} = 0, \label{Eq: a_exp_val} \\
	\mean{\hat{a}^2}_{\text{BSV}}^{(\text{APP})} &= \mean{(\hat{a}^\dagger)^2}_{\text{BSV}}^{(\text{APP})} = -\tanh(r) \cosh^2(r). \label{Eq: a_square_exp_val}
\end{align}

Inserting Eqs. (\ref{Eq: a_exp_val}) and (\ref{Eq: a_square_exp_val}) into Eq. (\ref{variance_general}), we find that 

\begin{align}
	\mean{\Delta\hat{X}^2(\theta)} &= 
	\frac{1}{2} \cosh^2 r \left[
	1 - \cos\!\left(2\theta\right)\tanh r
	\right]
	+ \frac{1}{4}.
\end{align}
To minimize this, we can take the derivative with respect to $\theta$, which vanishes for $\theta = 0,\pi/2$. Obviously, of these two, $\theta = 0$ minimizes the expression, and hence
\begin{align}
	\vartheta_{\text{BSV}}^{(\text{APP})} =  \min_{\theta\in[0,\pi)}\mean{\Delta\hat{X}^2(\theta)}
	=
	\frac{1}{2} \cosh^2 r \left(
	1 - \tanh r
	\right)
	+ \frac{1}{4}.
	\label{min_variance}
\end{align}
Notice especially that $\abs{\tanh x} \leq 1$ for all $x\in\mathbb{R}$, so $1 - \tanh r \geq 0$, which means that in the APP representation
\begin{align}
	\vartheta_{\text{BSV}}^{(\text{APP})} =  \min_{\theta\in[0,\pi)}\mean{\Delta\hat{X}^2(\theta)} \geq \frac{1}{4},
\end{align}
which does clearly does not match exact results where $\vartheta_{\text{BSV}}^{\text{Exact}} = \dfrac{1}{4} e^{-2r}$ \cite{Gerry2004}. 

Hence, the APP representation does not capture the correct photon statistics for Fock states or correct squeezing BSV fields and hence we abstain from considering these observables for the emitted field.

\section{Details on the lowest-order expansion of the Bessel functions}\label{App: Bessel_approx}
    For the discussion of the range of the perturbative regime of HHG in Sec. \ref{Sec: Power_scaling}, we are interested in determining the error in approximating the Bessel functions $J_n(x)$ to the lowest order. To this end, we employ Taylor's remainder theorem to obtain an upper bound on the error for some interval of $x$. 
    \par
    Specifically, we use the theorem that states that for an analytic function $f:[a,b]\in\mathbb{R}\rightarrow\mathbb{R}$ the remainder $R_n(x)$ of the $n$'th order Taylor expansion of $f$ can be estimated as for $x\in[a,b]$ as 
    \begin{align}
        \abs{R_n(x)}\leq M \frac{x^{n+1}}{(n+1)!},
    \end{align}
    where $M$ is a positive real number that fulfills 
    \begin{align}
        M\geq \abs{\frac{d^{n+1}}{dx^{n+1}}f(x)}, \quad \forall x\in [a,b].
    \end{align}
    \par
    Writing the Bessel function as a Taylor expansion we have
    \begin{align}
        J_n(x) = \sum_{k=0}^\infty \frac{(-1)^k}{k!(n+k)!}
        \left(
        \frac{x}{2}
        \right)^{n+2k},
    \end{align}
    which means that the lowest nonvanishing order in the expansion is $n$. Since we know that Bessel functions vanish when the argument is smaller than the order, we are motivated to consider the interval $[0,n/r]$, where $r$ is some positive real number that we will tune to give an acceptably low remainder.
    We fix this tolerance for the remainder to be $\abs{R_n(x)}\leq 1/100$, as we deem a 1\% error acceptable.
    \par 
    We can then determine the $M$ parameter for a given $r$ as
    \begin{align}
        M\geq \abs{\frac{d^{n+1}}{dx^{n+1}}J_n(x)}, \quad \forall x\in [0,\frac{n}{r}].
    \end{align}
    Using the identities for differentiating a Bessel function we may write
    \begin{align}
        \frac{d^{n+1}}{dx^{n+1}}J_n(x) = 
        \frac{1}{2^{n+1}}\sum_{k=0}^{n+1}\binom{n+1}{k}J_{2k-1}(x),
    \end{align}
    and utilizing that $\abs{J_n(x)}\leq 1/\sqrt{2}$ for all x and for $n\geq 1$, we conclude that
    \begin{align}
        \abs{\frac{d^{n+1}}{dx^{n+1}}J_n(x)}\leq \frac{1}{2^{n+1}}\sum_{k=0}^{n+1}\binom{n+1}{k} \frac{1}{\sqrt{2}}
        = \frac{1}{\sqrt{2}}.
    \end{align}
    Hence, we choose $M=1/\sqrt{2}$ to obtain that
    \begin{align}
        \abs{R_n(x)} &\leq \frac{1}{\sqrt{2}}\frac{x^{n+1}}{(n+1)!}
        \notag
        \\
        &\leq
        \frac{1}{\sqrt{2}}\frac{n^{n+1}}{r^{n+1}(n+1)!}, \label{Eq: restledsvurdering_1}
    \end{align}
    which is independent of $x$. Lastly, we want a universal estimate for all $n$. For this, we can use that $n! \geq \sqrt{2\pi} e(n/e)^n$, where $e$ is Euler's number. Hereby we can estimate that
    \begin{align}
        \abs{R_n(x)} &\leq \frac{1}{2\sqrt{\pi}e} \left( \frac{n}{n+1} \right)^{n+1} \left(\frac{e}{r}\right)^{n+1} 
        \notag
        \\
        &\leq
        \frac{1}{2\sqrt{\pi}e} \left(\frac{e}{r}\right)^{n+1},
    \end{align}
    which is a decreasing function of $n$ when $r \geq e$. We therefore find that
    \begin{align}
        \abs{R_n(x)} 
        \leq
        \frac{1}{2\sqrt{\pi}e} \left(\frac{e}{r}\right)^2,
    \end{align}
    for all $n\geq 1$. We can then solve the equation for $r$ given the tolerance
    \begin{align}
        \frac{1}{2\sqrt{\pi}e} \left(\frac{e}{r}\right)^2 &\leq \frac{1}{100}
        \notag
        \\
        \Rightarrow r &\geq 10\frac{\sqrt{e}}{2\sqrt[4]{\pi}} \approx 9.
        \label{Eq: r_vurdering}
    \end{align}
    As such we have found an estimate of an interval $[0,n/9]$, where the error of the Bessel functions is within the set tolerance. 
    \par 
    We note that this is a rough estimate since we have made an estimation that is valid for all $n$. For more precise results, which could yield lower $r$ values, one could consider each order separately and obtain an order dependent $r_n$. This could be done from Eq. (\ref{Eq: restledsvurdering_1}), and could also be improved by making a less rough estimation of the bounds of $J_n(x)$, which become smaller with the harmonic orders. The point is that we now know that the lowest-order approximation of $J_n(x)$ is valid within the tolerance $1/100$ on at least the interval $[0,n/9]$.
    \par 
    Hereby, in relation to Sec. \ref{Sec: Power_scaling}, we can say that the lowest-order approximation is valid for the Bessel function $J_n(l\tilde{g}_0 \abs{\alpha})$ when 
    \begin{align}
        \abs{\alpha} \leq \frac{n}{9l\tilde{g}_0} . 
    \end{align}

 We can then argue similarly to Sec. \ref{Sec: cutoff}, that we can use this approximation when the $P$ distribution is contained within this region. Again, we estimate the region where $P(\alpha)$ is not negligible as $\mu_P - 3\sigma_P \leq \abs{\alpha} \leq \mu_P + 3\sigma_P$. Hence, we can use the power-scaling law [Eq. (\ref{Eq: Power_scaling})] for the $n$'th harmonic when 
\begin{align}
	\mu_P + 3\sigma_P \leq n/(9l_{max}\tilde{g}_0).
	\label{Eq: perturbative_cutoff}
\end{align}
The parameters $\mu_P $ and $\sigma_P$ are the mean and standard derivation of the distribution $P(\abs{\alpha})$, respectively, which depends on the mean photon count of the driving field. Explicitly, they are given as
    \begin{align}
	\mu_P &= \int d^2\alpha P(\alpha) \abs{\alpha}
	\label{Eq: mu_P}
	\\
	\sigma_P^2 &= \int d^2\alpha P(\alpha) \abs{\alpha}^2 - 
	\left[ \int d^2\alpha P(\alpha) \abs{\alpha} \right]^2 
	\notag \\
	&= \mean{a^\dag_{\vec{k}_0,\sigma_0}a_{\vec{k}_0,\sigma_0}} - \mu_P^2 
	. \label{Eq: sigma_P}
\end{align} 
For a coherent state [Eq. (\ref{Eq: P_coherent})], the mean photon number is $\mean{ \hat{a}^\dagger_{\vec{k}, \sigma} \hat{a}_{\vec{k}, \sigma}} = \abs{\alpha_{\vec{k}, \sigma}}^2$ while for BSV fields [Eq. (\ref{Eq: P_bsv})] it is $\mean{\hat{a}^\dagger_{\vec{k}, \sigma} \hat{a}_{\vec{k}, \sigma}}=\sinh^2(r)$. For Fock and thermal fields, the mean photon number enters the respective distributions explicitly [see Eqs. (\ref{Eq: P_fock}) and (\ref{Eq: P_thermal})]. 
The parameters $\mu_P$ and $\sigma_P$ are calculated numerically using Eqs. (\ref{Eq: mu_P}) and (\ref{Eq: sigma_P}) for each value of mean photon number in the driving field. Thus, the inequality of Eq. (\ref{Eq: perturbative_cutoff}) is checked for each of these values, and the cutoff is set at the lowest mean photon number for which the inequality is broken. These cutoffs are plotted in Fig. \ref{Fig: Powerscaling}.

\section{Derivation of the time-resolved electric field} \label{App: E_field}

\begin{figure}[t]
	\includegraphics[width=\linewidth]{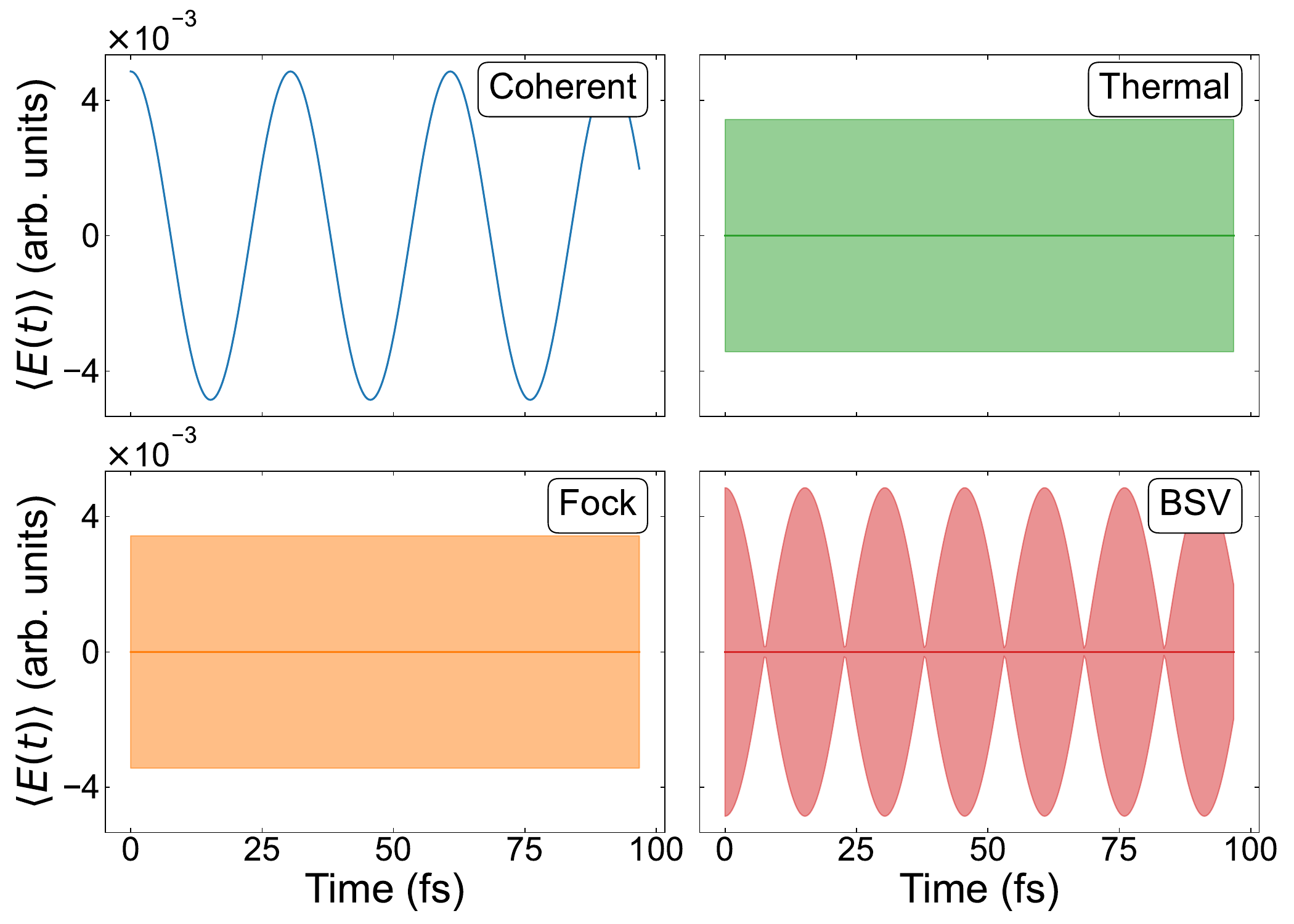}
	\caption{Time-resolved electric driving fields for coherent, Fock, thermal and BSV fields. Solid lines represent the expectation value of the electric field and the shaded area represent the uncertainty in the electric field neglecting the zero-point vacuum fluctuations. 
	}
	\label{Fig: E_field_drive_time}
\end{figure}
 In this appendix, we investigate the electric fields emitted from the HHG process. Before deriving the expressions for the emitted electric field, we first verify that the electric field of the driving laser is correctly reproduced in the APP. To this end, we compute $\mean{\hat{\vec{E}}(t)}$ with $\hat{\vec{E}}(t)$ given in Eq. (\ref{Eq: E_felt_definition}) for the driving fields using the $P$ functions  in Eqs. (\ref{Eq: P_coherent}), (\ref{Eq: P_thermal}), (\ref{Eq: P_fock}) and (\ref{Eq: P_bsv}). Likewise, we can compute the variance of the driving field and hereby the uncertainty in the field (neglecting zero-point fluctuations). These results are shown in Fig. \ref{Fig: E_field_drive_time}. It is seen that the fields are as expected: The coherent field has a well-defined frequency and amplitude with vanishing fluctuations. Thermal and Fock fields have vanishing mean fields, but their uncertainties are on the scale of the coherent state amplitude and are constant in time. The BSV field also has a vanishing mean field but the uncertainty oscillates in time with the same amplitude and twice the frequency of the coherent oscillations. As the APP representation reproduces the correct time-dependent driving fields for Fock and BSV, we trust its validity to produce the correct time-dependent generated fields.  

We now derive the equation for the mean and variance for the generated electric field. We consider the derived state of the field [Eq. (\ref{Eq: P_field})]
    \begin{align}
		\hat{\rho}_F(t) =
		\int d^2\alpha
		P(\alpha)
		\dyad{\gamma^\alpha_{k_0\sigma_0}(t)+\alpha}{\gamma^{\alpha}_{k_0\sigma_0}(t)+\alpha}
        \notag
        \\
        \bigotimes_{(k,\sigma)\neq (k_0,\sigma_0)}
		\dyad{\gamma^\alpha_{k\sigma}(t)}{\gamma^{\alpha}_{k\sigma}(t)},
		\label{Eq: P_field_appendix}
	\end{align}
    where 
    \begin{align}
		\gamma^\alpha_{k\sigma}(t)= 
		-i\frac{g_0}{\sqrt{\omega_k}} 
		\int_{t_i}^t \vec{j}^\alpha_{ii}(t')\cdot {\vec{e}}_\sigma e^{i\omega_k t'}dt'.
		\label{Eq: gamma_appendix}
	\end{align}
    The electric field operator in the dipole approximation is given again for clarity 
    \begin{align}
        \hat{\vec{E}}(t) = i\sum_{\vec{k}\sigma}g_0\sqrt{\omega_k}
        \left( 
            {\vec{e}}_\sigma \hat{a}_{\vec{k}\sigma}e^{-i\omega_k t} - h.c.
        \right).
    \end{align}
    Computing the expectation value of the electric field operator yields
    \begin{align}
        \mean{\hat{\vec{E}}(t)} = \int &d^2\alpha P(\alpha) \bigg[
        ig_0\sqrt{\omega_0}
        \left( 
            {\vec{e}}_{\sigma_0} (\gamma^\alpha_{k_0\sigma_0}(t)+\alpha)e^{-i\omega_0 t} - c.c.
        \right)
        \notag
        \\
        &+ 
        i\sum_{k\sigma\neq k_0\sigma_0}g_0\sqrt{\omega_k}
        \left( 
            {\vec{e}}_\sigma \gamma^\alpha_{k\sigma}(t)e^{-i\omega_k t} - c.c.
        \right)
        \bigg]
        \notag
        \\
        = 
        ig_0&\int d^2\alpha P(\alpha)
        \sum_{k\sigma}\sqrt{\omega_k}
        \left( 
            {\vec{e}}_\sigma \gamma^\alpha_{k\sigma}(t)e^{-i\omega_k t} - c.c.
        \right)
        \notag
        \\
        &+
        ig_0\sqrt{\omega_0}\int d^2\alpha P(\alpha)
        \left( 
            {\vec{e}}_{\sigma_0}\alpha e^{-i\omega_0 t} - c.c.
        \right).
    \end{align}
    The last term in this equation is just the expectation value of the driving field, which we shall henceforth neglect, since we are interested in the generated field. With this, we let $\sum_{\vec{k}} \mapsto  V/(2\pi c)^3 \int d\omega \omega^2\int d\Omega$, where $V$ is the quantization volume, $c$ the speed of light in vacuum and $d\Omega$ the solid angle infinitesimal. 

    \begin{align}
        \mean{\hat{\vec{E}}_{\text{HHG}}(t)} =& 
        i\frac{g_0 V}{(2\pi c)^3} 
        \int d^2\alpha P(\alpha) 
        \int_0^\infty d\omega\: \omega^2 \sqrt{\omega} 
        \notag
        \\
        & \quad \times \int d\Omega 
        \sum_\sigma 
        2i\textit{Im}\left(
        {\vec{e}}_\sigma\gamma^\alpha_{k\sigma}(t) e^{-i\omega t}
        \right)
        \notag
        \\
        =&
        \frac{2g_0^2 V}{(2\pi c)^3}
        \int d^2\alpha P(\alpha)  
        \textit{Im}\left[
        i  
        \int_{t_i}^t dt'
        \right. 
        \notag 
        \\
        & \left. \times
        \int d\Omega
        \sum_\sigma {\vec{e}}_\sigma
        \left(\vec{j}^\alpha_{ii}(t')\cdot {\vec{e}}_\sigma\right)
        \int_0^\infty d\omega\: \omega^2 
          e^{i\omega (t'-t)}
        \right].
    \end{align}
    We now seek to simplify this. Consider first the term
    \begin{align}
        \int d\Omega
        \sum_\sigma {\vec{e}}_\sigma
        \left(\vec{j}^\alpha_{ii}(t')\cdot {\vec{e}}_\sigma\right).
    \end{align}
	As a one-dimensional model of a solid is applied, we can let this direction be $\hat{\vec{z}}$ and take the polarization vectors be ${\vec{e}}_1 = \hat{\vec{\theta}}$ and ${\vec{e}}_2 = \hat{\vec{\phi}}$, which are the spherical coordinate unit vectors, that form an orthonormal basis of the orthogonal complement to the $k$ vector.
    Hereby $\vec{j}^\alpha_{ii}(t')\cdot {\vec{e}}_1 = -j^\alpha_{ii}(t')\sin\theta$ and $\vec{j}^\alpha_{ii}(t')\cdot {\vec{e}}_2 = 0$. Hereby it is found that $\sum_\sigma {\vec{e}}_\sigma
        \left(\vec{j}^\alpha_{ii}(t')\cdot {\vec{e}}_\sigma\right) = -\hat{\vec{\theta}}j^\alpha_{ii}(t')\sin\theta$, 
    and utilizing the spherical symmetry of the problem around the $\hat{\vec{z}}$ axis, it is found that
    \begin{align}
        \int d\Omega
        \sum_\sigma {\vec{e}}_\sigma
        \left(\vec{j}^\alpha_{ii}(t')\cdot {\vec{e}}_\sigma\right)
        &= -j^\alpha_{ii}(t') \hat{\vec{z}} \int_0^{2\pi} d\phi \int_0^{\pi} d\theta \sin^3\theta
        \notag 
        \\
        &= -\frac{8\pi}{3}\vec{j}^\alpha_{ii}(t'),
    \end{align}
    and we thereby get
    \begin{align}
        \mean{\hat{\vec{E}}_{\text{HHG}}(t)} &=
        \frac{2g_0^2 V}{3\pi^2 c^3}
        \int d^2\alpha P(\alpha)  
        \notag 
        \\ 
        \times &
        \text{Im}\left[
        i  
        \int_{t_i}^t dt'
        \vec{j}^\alpha_{ii}(t')
        \int_0^\infty d\omega\: \omega^2 
          e^{i\omega (t'-t)}
        \right].
    \end{align}
    Next, we rewrite the term
    \begin{align}
        &\text{Im}\left[
        -i  
        \int_{t_i}^t dt'
        \vec{j}^\alpha_{ii}(t')
        \int_0^\infty d\omega\: \omega^2 
          e^{i\omega (t'-t)}
        \right] 
        \notag
        \\
        &= 
        \text{Im}\left[
        i  
        \int_{t_i}^t dt'
        \vec{j}^\alpha_{ii}(t')
        \frac{d^2}{dt'^2 }
        \int_0^\infty d\omega
          e^{i\omega (t'-t)}
        \right]
        \notag
        \\
        &=
        \pi\int_{t_i}^t dt'
        \vec{j}^\alpha_{ii}(t')
        \frac{d^2}{dt'^2 }
        \delta(t'-t)
        \notag
        \\
        &=
        \pi
        \frac{d^2}{dt^2 }\vec{j}^\alpha_{ii}(t),
    \end{align}
    where we have utilized that $\vec{j}^\alpha_{ii}(t)$ is real, from being a diagonal matrix element, derivatives of distributions, and that $t\in[t_i,t]$. 
    \par 
    By writing out $g_0=\sqrt{2\pi/V}$, the front factor is then $\frac{4}{3c^3}$ and we achieve that the generated electric field is then given as
    \begin{align}
        \mean{\hat{\vec{E}}_{\text{HHG}}(t)} =
        -\frac{4}{3c^3} 
        \int d^2\alpha P(\alpha) \frac{d^2}{dt^2 }\vec{j}^\alpha_{ii}(t).
    \end{align}

The variance of the electric field can be derived in a similar way.

\newpage
\bibliography{bib_this}
	
\end{document}